\documentclass[twocolumn]{aastex61}

\usepackage{bm}

\newcommand\kms{\ifmmode{\rm km\thinspace s^{-1}}\else km\thinspace s$^{-1}$\fi}

\newcommand{\obj}{GW\,Ori}
\newcommand{\twelve}{${}^{12}$CO}
\newcommand{\thirteen}{${}^{13}$CO}
\newcommand{\eighteen}{C${}^{18}$O}

\received{October 9, 2017}
\revised{November 15, 2017}
\accepted{November 17, 2017}
\submitjournal{ApJ}

\begin{document}

\title{The Architecture of the GW~Ori Young Triple Star System and Its Disk: \\
Dynamical Masses, Mutual Inclinations, and Recurrent Eclipses}

\correspondingauthor{Ian Czekala}
\email{iczekala@stanford.edu}

\author[0000-0002-1483-8811]{Ian Czekala}
\altaffiliation{KIPAC Postdoctoral Fellow}
\affiliation{Kavli Institute for Particle Astrophysics and Cosmology,
Stanford University, 452 Lomita Mall, Stanford, CA 94305, USA}

\author[0000-0003-2253-2270]{Sean M. Andrews}
\affiliation{Harvard-Smithsonian Center for Astrophysics,
60 Garden Street, Cambridge, MA 02138, USA}

\author[0000-0002-5286-0251]{Guillermo Torres}
\affiliation{Harvard-Smithsonian Center for Astrophysics,
60 Garden Street, Cambridge, MA 02138, USA}

\author[0000-0001-8812-0565]{Joseph E. Rodriguez}
\affiliation{Harvard-Smithsonian Center for Astrophysics,
60 Garden Street, Cambridge, MA 02138, USA}

\author[0000-0002-4625-7333]{Eric L. N. Jensen}
\affiliation{Department of Physics and Astronomy,
Swarthmore College, 500 College Avenue, Swarthmore, PA 19081, USA}

\author[0000-0002-3481-9052]{Keivan G. Stassun}
\affiliation{Department of Physics and Astronomy, Vanderbilt University, 6301 Stevenson Center, Nashville, TN 37235, USA}
\affiliation{Department of Physics, Fisk University, Nashville, TN 37208, USA}

\author[0000-0001-9911-7388]{David W. Latham}
\affiliation{Harvard-Smithsonian Center for Astrophysics,
60 Garden Street, Cambridge, MA 02138, USA}

\author[0000-0003-1526-7587]{David J. Wilner}
\affiliation{Harvard-Smithsonian Center for Astrophysics,
60 Garden Street, Cambridge, MA 02138, USA}

\author[0000-0002-4020-3457]{Michael A. Gully-Santiago}
\affiliation{NASA Ames Research Center, Moffett Field, CA 94035, USA}

\author[0000-0001-5707-8448]{Konstantin N. Grankin}
\affiliation{Crimean Astrophysical Observatory, pos. Nauchnyi, Crimea, 298409 Russia}

\author[0000-0003-2527-1598]{Michael B. Lund}
\affiliation{Department of Physics and Astronomy, Vanderbilt University, 6301 Stevenson Center, Nashville, TN 37235, USA}

\author{Rudolf B. Kuhn}
\affiliation{South African Astronomical Observatory, P.O. Box 9, Observatory 7935, South Africa}

\author{Daniel J. Stevens}
\affiliation{Department of Astronomy, The Ohio State University, Columbus, OH 43210, USA}

\author[0000-0001-5016-3359]{Robert J. Siverd}
\affiliation{Las Cumbres Observatory Global Telescope Network, 6740 Cortona Drive, Suite 102, Santa Barbara, CA 93117, USA}

\author[0000-0001-5160-4486]{David James}
\altaffiliation{Event Horizon Telescope}
\affiliation{Harvard-Smithsonian Center for Astrophysics,
60 Garden Street, Cambridge, MA 02138, USA}

\author{B. Scott Gaudi}
\affiliation{Department of Astronomy, The Ohio State University, Columbus, OH 43210, USA}

\author{Benjamin J. Shappee}
\altaffiliation{Hubble, Carnegie-Princeton Fellow}
\affiliation{The Observatories of the Carnegie Institution for Science, 813 Santa Barbara St., Pasadena, CA 91101}

\author{Thomas W.-S. Holoien}
\altaffiliation{Carnegie Fellow}
\affiliation{The Observatories of the Carnegie Institution for Science, 813 Santa Barbara St., Pasadena, CA 91101}

\begin{abstract}
We present spatially and spectrally resolved Atacama Large Millimeter/submillimeter Array (ALMA) observations of gas and dust orbiting the pre-main sequence hierarchical triple star system \obj. A forward-modeling of the \thirteen\ and \eighteen\ $J$=2--1 transitions permits a measurement of the total stellar mass in this system, $5.29 \pm 0.09\,M_\odot$, and the circum-triple disk inclination, $137.6 \pm 2.0\degr$. Optical spectra spanning a 35 year period were used to derive new radial velocities and, coupled with a spectroscopic disentangling technique, revealed that the A and B components of \obj\ form a double-lined spectroscopic binary with a $241.50\pm0.05$ day period; a tertiary companion orbits that inner pair with a $4218\pm50$ day period. Combining the results from the ALMA data and the optical spectra with three epochs of astrometry in the literature, we constrain the individual stellar masses in the system ($M_\mathrm{A} \approx 2.7\,M_\odot$, $M_\mathrm{B} \approx 1.7\,M_\odot$, $M_\mathrm{C} \approx 0.9\,M_\odot$) and find strong evidence that at least one (and likely both) stellar orbital planes are misaligned with the disk plane by as much as 45\degr. A $V$-band light curve spanning 30 years reveals several new $\sim$30~day eclipse events 0.1--0.7~mag in depth and a 0.2 mag sinusoidal oscillation that is clearly phased with the AB--C orbital period. Taken together, these features suggest that the A--B pair may be partially obscured by material in the inner disk as the pair approaches apoastron in the hierarchical orbit. Lastly, we conclude that stellar evolutionary models are consistent with our measurements of the masses and basic photospheric properties if the \obj\ system is $\sim$1\,Myr old.
\end{abstract}

\section{Introduction} \label{sec:intro}

Pre-main sequence (pre-MS) stars in multiple systems---for which it is possible to precisely measure their fundamental stellar properties through dynamical means---serve as touchstones for understanding the final stages of stellar formation and the conditions under which planetary systems are assembled. While recent decades have seen steady progress in understanding binary formation in general, lingering uncertainties still remain about the characteristics of young spectroscopic binaries and higher order systems \citep{duchene13}. 

\obj, a G-type star associated with the $\lambda$ Orionis OB star-forming complex \citep{dolan00,dolan01,dolan02}, was one of the first T Tauri stars to be revealed as a spectroscopic binary, with a period of \mbox{240 days} \citep{mathieu91}. Radial velocity (RV) monitoring hinted at the presence of a third body with a period of $\sim$10\,years; a tertiary was confirmed directly using infrared interferometry \citep{berger11}. 
Circumstellar material in the \obj\ system was first inferred from infrared excess emission \citep{mathieu91}; a subsequent detection of the dust continuum at submillimeter wavelengths suggested the disk was especially massive and must be circumbinary \citep[$M_{\rm disk} \gtrsim 0.1 M_\odot$;][]{mathieu95}. 

The disk material provided a natural explanation for the quasi-periodic optical dimming of \obj\ over $\sim$30\,day durations: the suspicion was that a disk around the secondary was eclipsing the primary, presuming a nearly edge-on viewing angle \citep{shevchenko92,shevchenko98}.
\citet{fang17} spatially resolved the disk material with Submillimeter Array (SMA) observations of the dust continuum and the line emission from CO isotopologues, demonstrating its large radial extent and therefore presumably circum-triple architecture.  However, they found the disk has an \emph{intermediate} inclination to the line of sight ($i_\mathrm{disk} \approx 35\degr$), in apparent conflict with the eclipse model. Indeed, this adds to a collection of indirect evidence for a more complicated geometry in the inner disk, including mid-infrared fluxes that vary on $\sim$year timescales \citep{fang14}, and CO rovibrational emission lines with multi-component profiles \citep[which requires a complicated geometry and/or temperature structure in the inner disk;][]{najita03}.

Beyond resolving outstanding questions about its architecture, the \obj\ system presents an excellent opportunity to obtain a precise dynamical mass measurement for an earlier type ($\sim$G8) star at a very young age. Precise dynamical masses are crucial for calibrating the photospheric predictions (e.g., $T_\mathrm{eff}$, $L$) of stellar evolutionary models, and the region of the Hertzsprung-Russell (HR) diagram occupied by \obj\ is particularly sparsely populated with benchmark systems \citep{stassun14}. Especially interesting is the fact that three different dynamical mass measurement techniques can be employed to study the \obj\ system: (1) RV monitoring, which constrains the mass ratios of the stars \citep{mathieu91,fang14}; (2) astrometric monitoring, which provides the inclinations of the orbits and, when coupled with RV measurements, can reveal the individual component masses \citep{berger11}; and (3) the disk-based dynamical mass technique \citep[e.g.,][]{simon00,simon17,rosenfeld12b,czekala15a,czekala16}, which measures the {\it total} stellar mass.

In this paper, we combine information from each of these techniques to better understand the fundamental properties and underlying physical architecture of the \obj\ system.  Section~\ref{sec:obs} describes new ALMA observations of the \obj\ disk, an updated analysis of 35 years' worth of optical spectroscopic RV monitoring, and an extensive decades-long optical photometric catalog.  Section~\ref{sec:analysis} describes our tomographic reconstruction of the disk velocity field, a new analysis of the RV data, their combination with literature-based astrometric constraints \citep{berger11}, an assessment of the system parameters and geometry, and the connections to the observed photometric variations. Section~\ref{sec:discussion} concludes by discussing the structure and orientation of the disk with respect to the orbital architecture of the triple system, and considers the \obj\ system in the context of other young multiple pre-MS systems.

\section{Observations and Data Reduction \label{sec:obs}}

\subsection{Millimeter Interferometry}

\begin{figure*}[ht!]
\begin{center}
  \includegraphics[width=\linewidth]{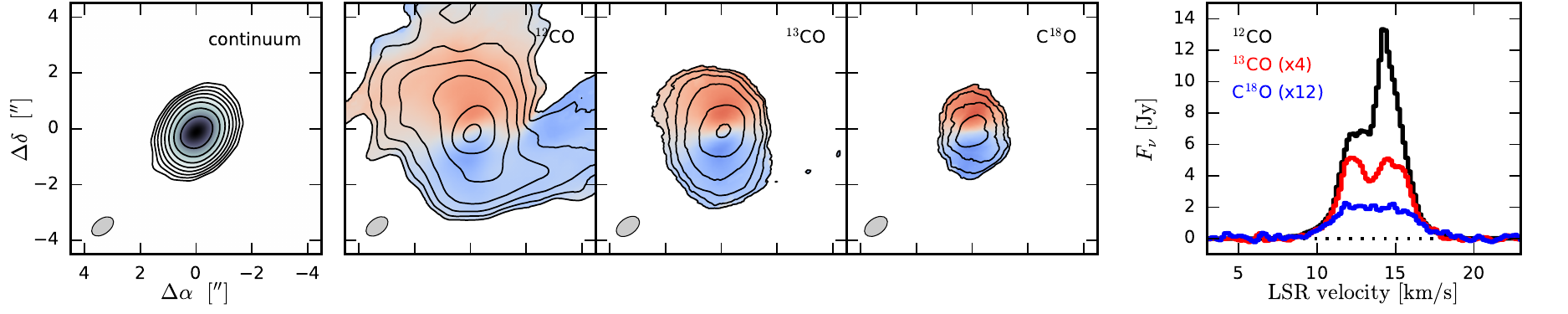}
  \figcaption{({\it left}) A 226\,GHz continuum image.  Contours start at 5$\times$ the RMS noise level and increase by factors of 2.  The synthesized beam geometry is shown in the lower left corner.  ({\it middle, left to right}) Maps of the $^{12}$CO, $^{13}$CO, and C$^{18}$O velocity-integrated intensities (contours, starting at 10, 3, and 3$\times$ the RMS noise levels, respectively, and increasing by factors of 2) overlaid on the intensity-weighted projected velocities (color-scale).  Note the prominent molecular cloud contamination in the $^{12}$CO map (see also Fig.~\ref{fig:chanmaps}).  ({\it right}) Spatially integrated spectra (inside the same {\tt CLEAN} mask, and smoothed with an 0.85\,km\,s$^{-1}$ Hanning kernel) for each CO line.
  \label{fig:moments}}
  \end{center}
\end{figure*}

GW Ori was observed with the ALMA interferometer on 2015 May 14 (program ID 2012.1.00496.S), with 37 of the 12\,m main array antennas configured to span baselines of 23--558\,m.  The double sideband Band 6 receivers were employed in dual polarization mode, and the ALMA correlator was set up to process data in 4 spectral windows (SPWs).  Two of these SPWs, centered at 220.426 and 230.450\,GHz to observe the $^{13}$CO and $^{12}$CO $J$=2--1 transitions (at rest frequencies of 220.399 and 230.538\,GHz, respectively), covered 234\,MHz of bandwidth in 3840 channels (a 61\,kHz channel spacing).  One other sampled 469\,MHz around 219.763\,GHz to observe the C$^{18}$O $J$=2--1 transition (at rest frequency 219.560\,GHz) with 3840 channels (a 122\,kHz channel spacing).  The last SPW sampled the continuum in a 1.875\,GHz range around 231.956\,GHz using 128 coarse channels (a 15.625\,MHz channel spacing).

The observations cycled between GW Ori and the quasar J0510+1800 with a 7 minute cadence.  The quasar J0423-0120 and Ganymede were observed as bandpass and flux calibration sources, respectively, at the start of the execution block.  The total on-source integration time for GW~Ori was 16 minutes.  The observing conditions were typical for Band 6 projects, with a precipitable water vapor level around 1.1\,mm.

The visibility data were calibrated with standard procedures using the {\sc CASA} software package (v4.4).  The raw, observed visibility phases were adjusted based on the contemporaneous measurements of water vapor radiometers, flagged when applicable, and then the bandpass shape in each SPW was calibrated based on the observations of J0423-0120.  The absolute amplitude scale was determined based on the observations of Ganymede.  The complex gain behavior of the array and atmosphere was corrected based on the repeated observations of J0510+1800.  The calibrated visibilities showed a strong continuum signal, suggesting that self-calibration could significantly improve the data quality.  An initial model based on a preliminary continuum image was used for two rounds of phase-only self-calibration (on 30\,s, then 6\,s intervals) and one additional round that included the amplitudes (on a 7 minute scan interval).  This self-calibration reduced the RMS noise level in the continuum by a factor of $\sim$40.  After applying the self-calibration tables to the entire dataset (channel by channel),  we parsed out data products for each individual emission tracer of interest.  A set of continuum visibilities was constructed by spectrally averaging the line-free channels in each SPW into $\sim$125\,MHz increments.  The spectral visibilities for the $^{12}$CO, $^{13}$CO, and C$^{18}$O lines were continuum-subtracted and regridded into 170\,m s$^{-1}$-wide channels in the LSRK restframe over a $\sim$10\,km s$^{-1}$ range around the line centers.

\begin{figure*}[ht!]
\begin{center}
  \includegraphics{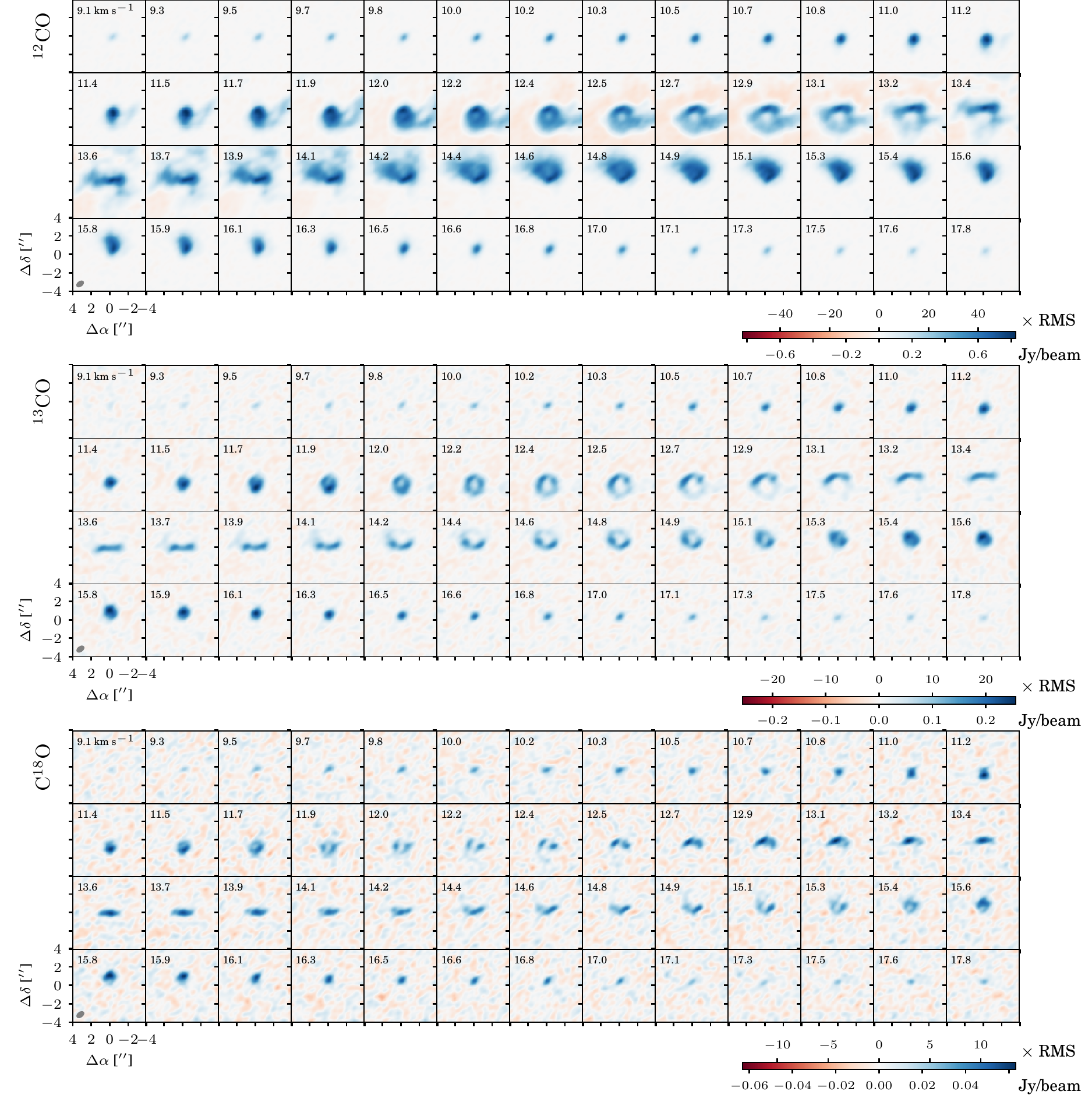}
  \figcaption{Channel maps of the $^{12}$CO, $^{13}$CO, and C$^{18}$O ({\it from top to bottom}) line emission from the GW~Ori disk.  Each channel represents the emission in a 170\,m\,s$^{-1}$-wide velocity bin.  LSRK velocities are indicated in the upper left, and synthesized beam sizes in the lower left of each panel.  Scale bars are provided at the bottom right of each set of channel maps.
  \label{fig:chanmaps}}
  \end{center}
\end{figure*}

These fully reduced visibility sets were then imaged by Fourier inversion assuming a Briggs (robust=0.5) weighting scheme and deconvolution with the standard {\tt CLEAN} algorithm.  Some basic image properties for the synthesized continuum image and spectral line image cubes are listed in Table~\ref{tab:ims}.  The continuum and spectral line moment maps are shown together in Figure~\ref{fig:moments}, along with a comparison of the integrated spectra.  The channel maps for individual lines are compiled in Figure~\ref{fig:chanmaps}.

\begin{deluxetable}{lcc}[b!]
\tablewidth{18pc}
\tablecaption{ALMA Image Properties  \label{tab:ims}}
\tablehead{
\colhead{} &
\colhead{} &
\colhead{RMS}
\\
\colhead{} &
\colhead{beam dimensions} &
\colhead{mJy beam$^{-1}$}}
\startdata
226\,GHz continuum  & $0\farcs88\times0\farcs54$, 126\degr\ & 0.055 \\
$^{12}$CO $J$=2$-$1 & $0\farcs89\times0\farcs56$, 126\degr\ & 6     \\
$^{13}$CO $J$=2$-$1 & $0\farcs93\times0\farcs59$, 126\degr\ & 8     \\
C$^{18}$O $J$=2$-$1 & $0\farcs92\times0\farcs58$, 126\degr\ & 5     \\
\enddata
\tablecomments{The RMS noise levels recorded for the spectral line cubes correspond to the values per 170\,m s$^{-1}$ channel.}
\end{deluxetable}

The 226\,GHz (1.3\,mm) continuum map shows a bright (\mbox{flux density = $202\pm20$\,mJy}), compact but marginally resolved (deconvolved Gaussian FWHM $\approx 0\farcs9$) source centered on the GW~Ori stellar system, with a peak intensity of 67\,mJy beam$^{-1}$ (S/N $\approx 1200$). Our integrated flux density measurement is consistent with that of \citet[$255 \pm 60$\,mJy]{mathieu95},  but marginally discrepant with that of \citet[$320 \pm 64$\,mJy]{fang17}. A crude estimate of the emission geometry (from a Gaussian fit to the visibilities) suggests an inclination of \mbox{35--40\degr}, with the major axis oriented $\sim$170\degr\ E of N.

The CO isotopologue channel maps reveal bright (integrated intensities of 41.8, 5.7, and 0.8\,Jy\,km\,s$^{-1}$ for $^{12}$CO, $^{13}$CO, and C$^{18}$O, respectively) and extended (FWHM $\sim 2\farcs5$) emission that is clearly in rotation around the continuum centroid, spanning a projected velocity range of $\pm$5\,km s$^{-1}$ from the line center.  The line emission is blueshifted to the south and redshifted to the north, consistent with the orientation estimated from the continuum emission. The peak intensities for each line are $\sim$800, 290, and 55\,mJy beam$^{-1}$ in the brightest channels (peak S/N $\approx 130$, 35, and 14) for \twelve, \thirteen, and \eighteen, respectively. The \twelve\ channel maps show some clear evidence for structured contamination from the surrounding molecular cloud, particularly as a streamer to the west at $\sim$11--13\,km s$^{-1}$ and some diffuse clumps to the north around \mbox{13--14\,km s$^{-1}$}, confirming the ``tail''-like feature seen by \citet{fang17}.  These structures are much fainter, but still present, in $^{13}$CO emission; they are not apparent in the C$^{18}$O maps.

\subsection{Optical Spectroscopy \label{subsec:spectroscopy}}

\obj\ was monitored spectroscopically at the Harvard-Smithsonian Center for Astrophysics for more than 35 years, beginning in 1981 November. A total of 203 usable spectra were gathered through 2009 April using three nearly identical echelle spectrographs (Digital Speedometers, DS; now decommissioned) with a resolving power of $R \approx 35,000$ mounted on three different telescopes: the 1.5\,m Tillinghast reflector at the Fred L.\ Whipple Observatory (Mount Hopkins, AZ), the 4.5\,m-equivalent Multiple Mirror Telescope (also on Mount Hopkins) before conversion to a monolithic mirror, and occasionally on the 1.5\,m Wyeth reflector at the Oak Ridge Observatory (in the town of Harvard, MA).  Each instrument was equipped with an intensified photon-counting Reticon detector limiting the output to a single echelle order 45~\AA\ wide, which was centered on the region of the \ion{Mg}{1}\,b triplet at 5187~\AA\ \citep[see][]{latham92}. The signal-to-noise ratios of these observations range from 14 to 59 per resolution element of 8.5~\kms, with a median of 41. Wavelength calibrations were based on exposures of a thorium-argon lamp taken before and after each science exposure. Reductions were performed with a dedicated pipeline, and the zero-point of the velocities was monitored regularly by means of exposures of the evening and morning twilight sky. The original analysis of \cite{mathieu91} used a subset of 45 of these spectra. A further 79 usable spectra of \obj\ were collected with the Tillinghast Reflector Echelle Spectrograph \citep[TRES;][]{furesz08}, a bench-mounted, fiber-fed echelle instrument attached to the 1.5\,m Tillinghast reflector. It provides a resolving power of $R \approx 44,000$, delivering 51 orders covering the wavelength interval 3900--9100~\AA. These observations were made between 2010 November and 2017 April.  Signal-to-noise ratios at 5200~\AA\ range from 28 to 195 per resolution element of 6.8~\kms, with a median of 74. Wavelength calibration was carried out as above, and reductions were performed as described by \cite{buchhave10}. RV standard stars were observed each night to monitor the zero point and place it on the same system as the DS observations to within $\sim$0.1~\kms.

All of our spectra appear to be single-lined, with broad features indicative of significant rotation. Preliminary RV measurements were therefore made with standard one-dimensional cross-correlation techniques, as in the analysis of \cite{mathieu91}. However, several pieces of evidence suggested it should be possible to detect the lines of the secondary in the 240 day binary. In particular, the large flux ratio of $f_B/f_A = 0.57 \pm 0.05$ (weighted average) reported by \cite{berger11} in the $H$-band, when translated to the optical, would still be significant for any reasonable assumption of the effective temperatures, making our non-detection of the secondary somewhat surprising. Furthermore, those same authors proposed that the system is observed nearly face-on, which would lead to strong line blending that could explain our lack of detection despite the sizable brightness of the secondary. 

We embarked on a search for such a  signature using the then-under-development \texttt{PSOAP} spectroscopic disentangling package \citep{czekala17}, with the assumption that it must be at or near the detection limit (e.g., $q_\mathrm{in} \lesssim 0.2$) since it had not been previously seen.
Given a time-series of high resolution spectroscopic observations covering the orbital phase of the binary or triple star, \texttt{PSOAP} simultaneously infers the intrinsic spectrum of each star along with the stellar orbit using Gaussian processes as a modeling basis. This provides a robust probabilistic inference of both the orbits and spectra in a purely data-driven manner, which can further be used to measure fundamental properties with traditional analysis techniques.
Preliminary results hinted at the detection of the secondary, but for mass ratios much larger than expected ($q_\mathrm{in} > 0.5$). Because the algorithm was not yet fully vetted, we discounted those results. To our excitement, however, shortly thereafter we learned that \obj\ had been revealed as a \emph{double}-lined binary based on high resolution infrared spectroscopy ($q_\mathrm{in} \sim 0.55$, Prato et al. (2017), \emph{submitted}). Motivated by that result, we renewed our efforts to search for the secondary using \texttt{PSOAP} and a targeted \texttt{TODCOR} analysis, which is a two-dimensional cross-correlation algorithm designed to minimize biases in the RVs due to line blending.

At present, one limitation of the \texttt{PSOAP} framework (and Gaussian processes in general, to some extent) is the extreme computational expense in performing large matrix calculations. This generally limits us to considering fewer than 20 epochs of high resolution spectra at a time, which consequentially limits the complexity of the orbital model that can be used. Although it was straightforward to extend the framework to utilize a hierarchical triple orbital model and three Gaussian process components, we found that we were unable to employ enough spectroscopic epochs to sufficiently constrain the more complex orbital model. Therefore, we experimented using different subsets of the highest S/N data in the range of 5060 - 5290\AA\ to test our sensitivity to the presence of the secondary and tertiary spectral signatures. In all of these tests, we clearly detected the features of the secondary but found no obvious evidence for spectroscopic signatures of the tertiary.

\begin{deluxetable}{lcccc}[t!]
\tablecaption{Heliocentric RV measurements of \obj.
\label{tab:RVs}}
\tablehead{
\colhead{HJD} &
\colhead{$RV_{\rm A}$} &
\colhead{$\sigma_{\rm A}$} &
\colhead{$RV_{\rm B}$} &
\colhead{$\sigma_{\rm B}$}
\\
\colhead{[2,400,000$+$]} &
\colhead{[\kms]} &
\colhead{[\kms]} &
\colhead{[\kms]} &
\colhead{[\kms]}}
\startdata
44919.0042 &  31.24  &  5.40 &  28.50 &  19.13 \\
45301.8865 &  25.10  &  5.18 &  25.29 &  18.35 \\
45336.7941 &  23.46  &  3.36 &  20.37 &  11.92 \\
45708.7038 &  33.05  &  5.83 &  20.37 &  20.65 \\
45709.6058 &  37.70  &  2.76 &  25.18 &   9.77 \\
\enddata
\tablecomments{Observations up to HJD 2,454,926.6573 were obtained
  with the DS, and the remainder with TRES. This table is available in its entirety in machine-readable
  form.}
\end{deluxetable}

Based on the guidance from these results, we re-examined our spectra with {\tt TODCOR} and succeeded in detecting the secondary via cross-correlation, as well. As anticipated, the lines of the two stars are always heavily blended, which causes a strong degeneracy between the adopted rotational line broadening for the templates (see below), the velocity amplitudes, the adopted temperatures, and the flux ratio. To measure RVs we adopted synthetic templates from the PHOENIX library of \cite{husser13}, broadened to match the resolution of our spectra. For the TRES observations we restricted our analysis to the 100~\AA\ order centered on the \ion{Mg}{1}\,b triplet, both for consistency with the analysis of the DS spectra, which cover only a 45~\AA\ window centered on this region, and because experience shows that it contains most of the information on the velocities. The one-dimensional cross-correlations needed to construct the 2-D correlation function in {\tt TODCOR} were computed using the IRAF\footnote{IRAF is distributed by the National Optical Astronomy Observatories, which are operated by the Association of Universities for Research in Astronomy, Inc., under cooperative agreement with the National Science Foundation.} task {\tt XCSAO} \citep{kurtz98}. The template parameters were selected based on an analysis of the stronger TRES spectra, as follows. For the primary star we adopted a temperature of $T_{\rm eff} = 5700$~K proposed by \cite{mathieu91}, along with $\log g = 3.0$ and solar metallicity, although the latter has minimal effect. The same composition and surface gravity were used for the secondary. The rotational broadening ($v \sin i$) of the primary, the secondary temperature, and the secondary $v \sin i$ were then determined by running extensive grids of 2-D cross-correlations over broad ranges in each parameter in a manner similar to that described by \cite{torres02}, seeking the best match between the templates and the observed spectra as measured by the peak cross-correlation coefficient averaged over all exposures. For each combination of template parameters we also determined the flux ratio that maximizes the correlation.

In this way we determined a best-fit secondary temperature of $T_{\rm eff} = 4800 \pm 200$~K, and $v \sin i$ values for the primary and secondary of 40 and 45~\kms, respectively, with estimated uncertainties of 5~\kms. The measured flux ratio in the \ion{Mg}{1}\,b 5187~\AA\ region is $f_B/f_A = 0.25 \pm 0.05$. While in principle these temperatures and $v \sin i$ values are merely free parameters that provide the best match to the observed spectra, in the following we interpret them also as estimates of the physical properties of the stars. The RVs we measured from our DS and TRES spectra with these parameters are reported in Table~\ref{tab:RVs}, along with their uncertainties.
Typical errors for the primary and secondary are 1.0 and 2.7~\kms\ for TRES, and 2.5 and 8.7~\kms\ for the DS measurements, though individual errors can sometimes be much larger. Despite the use of {\tt TODCOR}, we reiterate that the severe line blending at all phases of the inner orbit caused by a combination of rotational broadening and small velocity amplitudes makes the RVs very susceptible to errors in the template parameters (particularly $v \sin i$) and in the adopted flux ratio, and as a result the orbital elements presented later may suffer from systematic errors not included in the statistical uncertainties. Nevertheless, as a consistency check we used PHOENIX spectra from \cite{husser13} for the primary and secondary stellar parameters given above to extrapolate our measured flux ratio at 5187~\AA\ to the near infrared, and obtained an $H$-band value of $f_B/f_A = 0.57 \pm 0.12$. While less precise than the \cite{berger11} measurement, the agreement is excellent.

\subsection{Time-series Photometry}

We have assembled a high cadence lightcurve of \obj\ covering a $\sim$30 year timespan by drawing from several ongoing photometric surveys as well as archival observations. The details of the different surveys used in compiling this catalog are described below.

\subsubsection{Maidanak Observatory}

Photoelectric $UBVR$ observations of \obj\ were obtained at Mount Maidanak Observatory in Uzbekistan. About 530 $UBVR$ magnitudes were collected from 1987 to 2003, although the number of $U$ measurements is relatively small compared to the other photometric bands.
All observations were performed with three telescopes (one 0.48\,m and two 0.6\,m reflectors) using identical single-channel pulse-counting photometers with FEU-79 photomultiplier tubes. The observations of \obj\ were carried out as part of the ROTOR program, which was described by \citet{shevchenko93}.

The RMS uncertainty for a single measurement in the instrumental system was $0.01$\,mag in $BVR$ and $0.02$\,mag in $U$. Observations were carried out either differentially using a nearby reference star or directly by estimating the nightly extinction. In the latter case, several reference stars were observed every night to derive the extinction coefficients in each filter. Selected standard stars were observed and used to calibrate instrumental magnitudes on the Cousins system. We then transformed the magnitudes to the Johnson
$UBVR$ system using the relationship: $(V-R)_\mathrm{C} = -0.0320 + 0.71652(V-R)_\mathrm{J}$. The systematic uncertainty in this conversion is $0.01\,$mag. 

\subsubsection{KELT}

The Kilodegree Extremely Little Telescope (KELT) project uses two telescopes to survey over 70\% of the entire sky searching for transiting planets around bright stars ($8<V<11$). The telescopes, located in Sonoita, AZ (KELT-North) and Sutherland, South Africa (KELT-South), have a 42\,mm Mamiya 645-series wide-angle lens resulting in a $26^{\circ}$ $\times$ $26^{\circ}$ field-of-view (FOV), and a 23$\arcsec$ pixel scale. Both telescopes use a broad $R$-band filter. KELT observes using a Paramount ME German equatorial mount with a 180$^{\circ}$ meridian flip; therefore KELT observes in either an ``east'' or ``west'' orientation. The telescope optics are not perfectly axisymmetric, and so the point spread function (PSF) changes from one orientation to the other. Throughout the data reduction process, the east and west observations are treated as though they were acquired from separate telescopes. For \obj\ specifically, the PSF asymmetry results in a 0.2\,mag systematic offset between the east and west orientations. See \citet{Siverd12} and \citet{Kuhn16} for a detailed description of the KELT observing strategy and reduction process. \obj\ was located in KELT-South field 05 ($\alpha$ =  06hr 07m 48.0s, $\delta$ = $+3^{\circ}$ 00$\arcmin$ 00$\arcsec$) and was observed 2889 times from UT 2010 February 28 until UT 2015 April 09, with a median uncertainty of 0.005 mag.

\subsubsection{ASAS}

Using two observing locations, in Las Campanas, Chile and Haleakala, Maui, the All-Sky Automated Survey (ASAS) project was designed to observe the entire sky to a limiting optical magnitude of 14. The two observatory setups each contained a wide-field Minolta 200/2.8 APO-G telephoto lenses with a 2K$\times$2K Apogee CCD and both observed simultaneously in $B$ and $V$ band. The telescope and camera set up correspond to a $8\fdg8\times8\fdg8$ field-of-view. ASAS observed GW Ori in the $V$ band from UT 2001 March 11 until UT 2009 November 29, obtaining 480 observations with a median per-point uncertainty of 0.036 mag.

\subsubsection{ASAS-SN}

Focused on the discovery and characterization of supernovae, the All-Sky Automated Survey for SuperNovae \citep[ASAS-SN;][]{Shappee14, kochanek17} surveys the entire sky down to $V \sim 17$ mag every \mbox{$\sim$2 days}. Hosted by the Las Cumbres Observatory (LCO) at Mount Haleakala, Hawaii and the Cerro Tololo InterAmerican Observatory (CTIO) in Chile, each location hosts four 14\,cm Nikon telephoto lenses with a 2k $\times$ 2k thinned CCD \citep{Brown13}. The telescopes have a $4\fdg5\times4\fdg5$ field-of-view and a 7$\farcs$8 pixel scale. ASAS-SN obtained 799 observations of GW Ori from UT 2014 December 16 until UT 2017 March 15, with a typical per point error of 0.01 mag. Table~\ref{tab:phot} lists all of the photometric observations from the aforementioned telescopes.

\begin{deluxetable}{lcccc}[t!]
\tablecaption{Photometric measurements of \obj.
\label{tab:phot}}
\tablehead{
\colhead{HJD} &
\colhead{$m_V$} &
\colhead{$\sigma_V$} & 
\colhead{Telescope}
\\
\colhead{[2,400,000$+$]} &
\colhead{[mag]} &
\colhead{[mag]} &
}
\startdata
47031.4670 & 9.94 & \nodata &  Maidanak \\
47032.4760 & 9.90 & \nodata &  Maidanak \\
47034.4826 & 9.86 & \nodata &  Maidanak \\
47035.4806 & 9.88 & \nodata &  Maidanak \\
47036.4839 & 9.87 & \nodata &  Maidanak \\
\enddata
\tablecomments{This table is available in its entirety in machine-readable form.}
\end{deluxetable}

\section{Analysis and Results \label{sec:analysis}}
These new datasets allow us to study the architecture of the \obj\ system in a comprehensive manner. First, we forward-model the ALMA \thirteen\ and \eighteen\ transitions to reconstruct the disk velocity field, measure its inclination, and determine the total stellar mass of \obj. Next, we fit a hierarchical triple star model to the RVs and archival astrometry to determine individual stellar masses and orbital inclinations, and compare these properties to that of the disk. Last, we use the extensive lightcurve of \obj\ to identify new eclipse events and oscillatory modes, and compare these to the orbital periods to inform theories of their physical origin.

\subsection{A Reconstruction of the Disk Velocity Field \label{sec:disk}}

We use the spatially and spectrally resolved molecular line emission observed with ALMA to tomographically reconstruct the disk velocity field and make a dynamical measurement of the total stellar mass in the \obj\ system. We follow the forward modeling procedures described by \citet{czekala15a,czekala16} using the associated open-source software package {\tt DiskJockey}.\footnote{Available under an MIT license at \url{https://github.com/iancze/DiskJockey}.}

\begin{figure*}[ht!]
\begin{center}
  \includegraphics{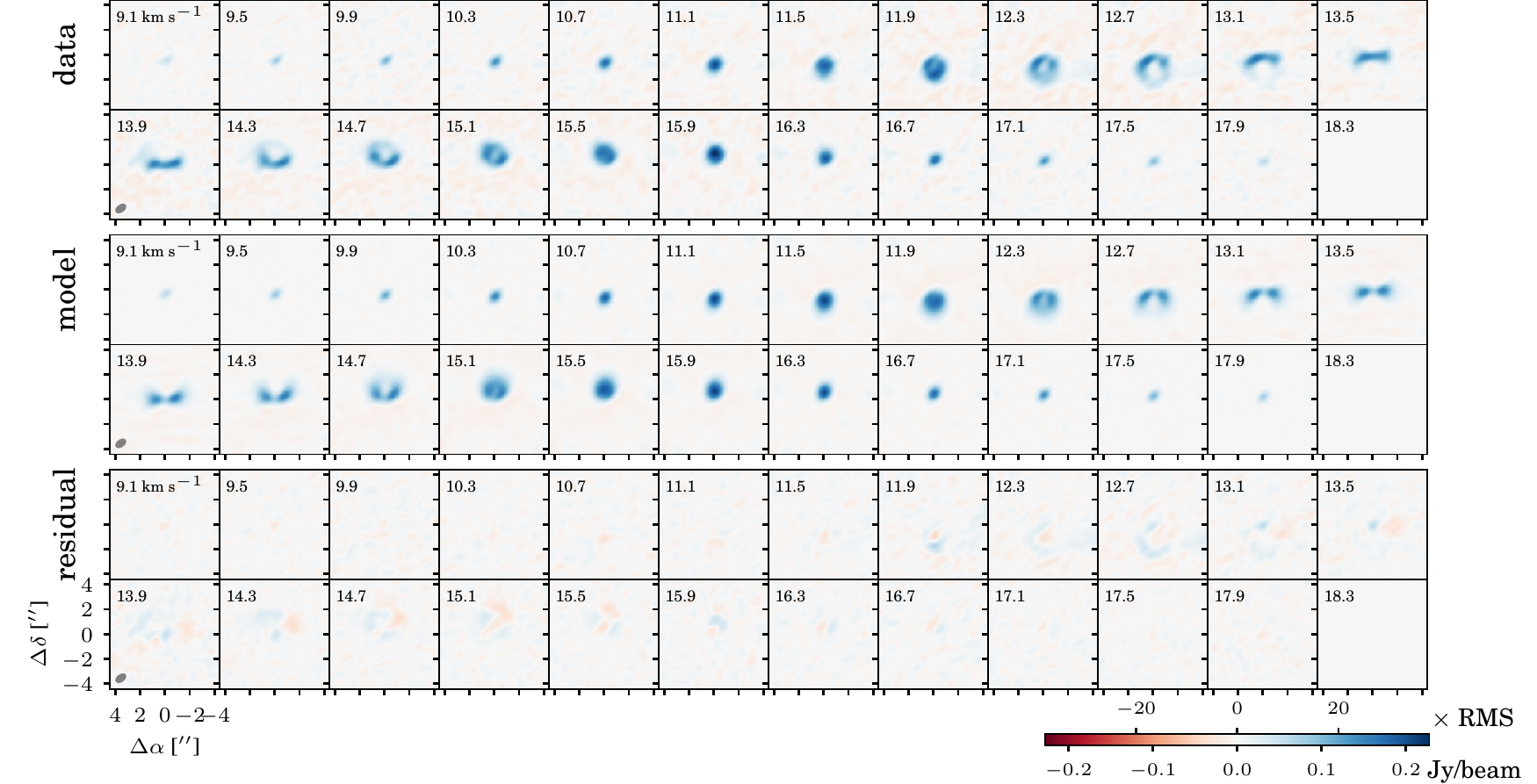}
  \figcaption{A comparison of the observed channel maps of the $^{13}$CO line emission ({\it top}) with a best-fit model ({\it middle}; constructed from a synthetic visibility set based on the inferred parameters listed in Table~\ref{table:components} and then imaged in the same way as the data) and the associated residuals ({\it bottom}; the imaged data$-$model residual visibilities).  The annotation is the same as in Fig.~\ref{fig:chanmaps}.
  \label{fig:13co}}
  \end{center}
\end{figure*}

The basis of the parametric physical model adopted in this approach is a radial surface density profile, $\Sigma(r)$, designed to mimic a simple theoretical description for a viscous accretion disk \citep{lyndenbell74,hartmann98}.  That profile decreases like $1/r$ interior to a characteristic radius $R_c$, and has an exponential taper $e^{-r/R_c}$ at larger radii.  The vertical distribution of density is controlled by the temperature structure: we assume a vertically isothermal model with a radial profile $T(r) = T_{10} (r/10$\,AU$)^{-q}$. To convert the total gas densities to the volume density of a given species, we use abundance ratios that are representative of the dense interstellar medium: [H$_2$/ gas] = 0.8, [H/H$_2$] = 2, [$^{12}$CO/H] = $7.5 \times 10^{-5}$, [$^{12}$CO/$^{13}$CO] = 69, and [$^{12}$CO/C$^{18}$O] = 557 \citep[e.g.,][]{henkel94,prantzos96}.

\begin{figure*}[ht!]
\begin{center}
  \includegraphics{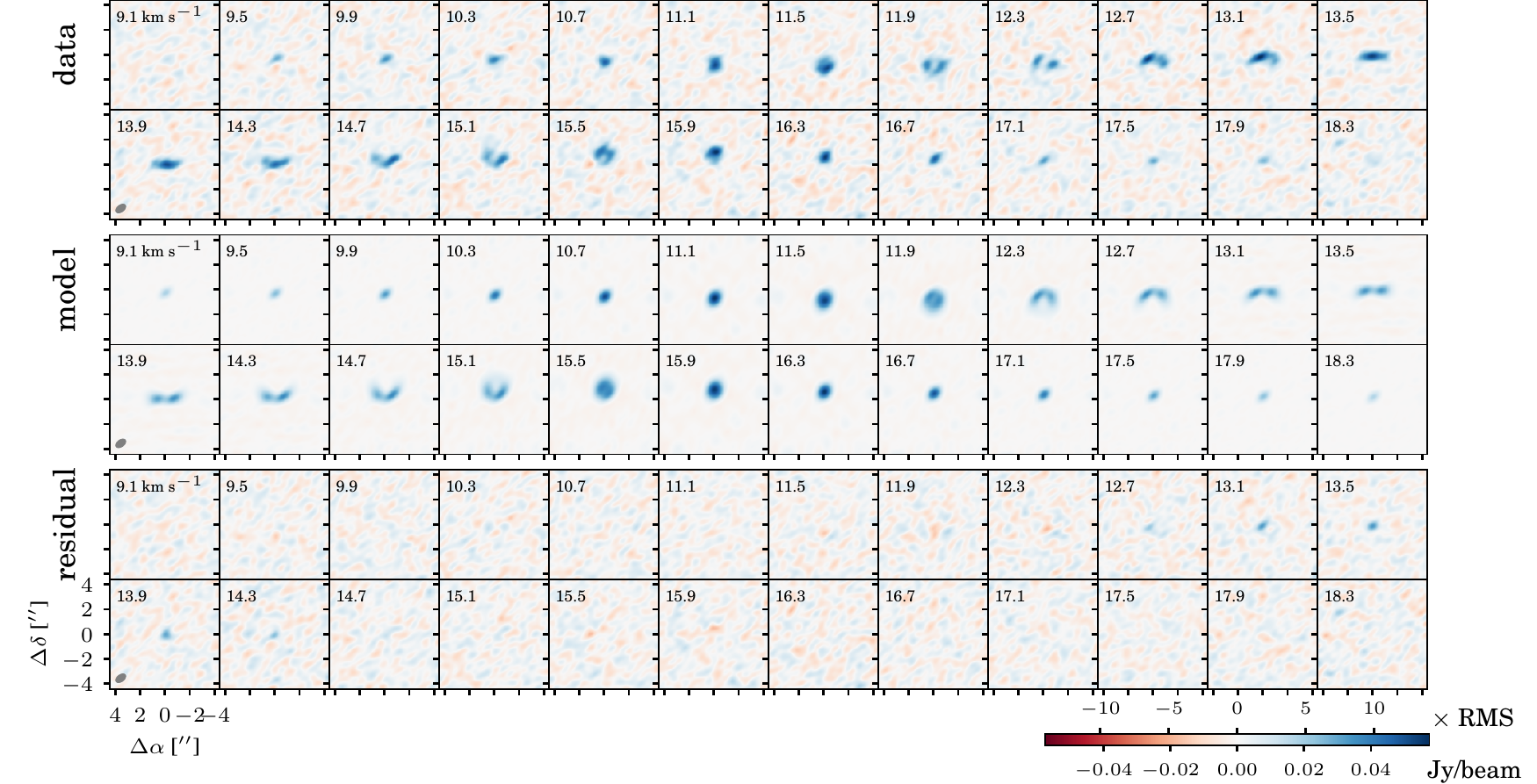}
  \figcaption{A comparison of the observed channel maps of the C$^{18}$O line emission ({\it top}) with a best-fit model ({\it middle}; constructed from a synthetic visibility set based on the inferred parameters listed in Table~\ref{table:components} and then imaged in the same way as the data) and the associated residuals ({\it bottom}; the imaged data$-$model residual visibilities).  The annotation is the same as in Fig.~\ref{fig:chanmaps}.
  \label{fig:c18o}}
  \end{center}
\end{figure*}

The disk kinematics are assumed to be Keplerian and dominated by the total stellar mass $M_\mathrm{tot}$, with a velocity field that appropriately accounts for the two-dimensional distribution of the emitting layer \citep[see][]{rosenfeld13a}.  The  line-spread function is characterized with a width defined by the quadrature sum of thermal and non-thermal ($\xi$; presumably turbulent) contributions. For any physical structure specified by these 6 parameters, \{$\Sigma_c$, $R_c$, $T_{10}$, $q$, $M_\mathrm{tot}$, $\xi$\}, we solve the molecular rate equations (assuming LTE) and ray-trace the associated emission into a set of high resolution channel maps using the radiative transfer package {\tt RADMC-3D} \citep{dullemond12}. That ray-tracing requires that we specify 5 additional geometric parameters: the disk inclination to the line-of-sight ($i_\mathrm{disk}$), the position angle of the disk rotation axis projected on the sky ($\varphi$), the LSRK systemic radial velocity ($v_r$), and a pair of positional offsets from the observed pointing ($\mu_\alpha$, $\mu_\delta$). We adopt a fixed distance to the \obj\ system, $d = 388\,$pc \citep{kounkel17}, to make the problem more computationally tractable; the effects of this assumption are discussed in Section~\ref{sec:joint}. The GAIA DR1 parallax to \obj\ is still rather uncertain: the mean estimate is slightly larger than our adopted distance but its large uncertainty means that it is still consistent at the $1\,\sigma$ level \citep[$\pi = 2.1 \pm0.5\,$mas;][]{gaia16}.

The model channel maps are then Fourier transformed and sampled at the same spatial frequencies observed by ALMA. The model quality with respect to the observed visibilities is evaluated with a $\chi^2$ likelihood function that incorporates the nominal visibility weights. We assume flat priors on all parameters except for $i_\mathrm{disk}$, where instead we adopt a simple geometric prior \citep[the disk angular momentum vector is distributed uniformly on a sphere; e.g.,][]{czekala16}. The posterior distribution of these parameters is explored using Markov Chain Monte Carlo (MCMC) simulations with the affine invariant ensemble sampler proposed by \citet{goodman10}, as implemented in the {\tt emcee} code \citep{foreman-mackey13} and ported to the {\tt Julia} programming language (included in {\tt DiskJockey}).

\begin{deluxetable}{lcc}
\tablecaption{Inferred Disk Model Parameters\label{table:components}}
\tablehead{\colhead{Parameter} & \thirteen & \eighteen}
\startdata
$M_\mathrm{tot}\quad [M_\odot]$ & $5.28 \pm 0.06$ & $5.38 \pm 0.23$ \\
$r_c$ [au] & $237 \pm 5$ & $151 \pm 21$ \\
$T_{10}$ [K] & $51 \pm 2$ & $32 \pm 4$ \\
$q$ & $0.308 \pm 0.012$ & $0.378 \pm 0.037$ \\
$\log_{10} M_\mathrm{disk} \quad \log_{10} [M_\odot]$ & $-1.69 \pm 0.02$ & $-1.02 \pm 0.22$ \\
$\xi\,[\kms]$ & $0.59 \pm 0.01$ & $0.37 \pm 0.03$ \\
$i_d \quad$ [deg] & $137.7 \pm 0.3$ & $135.2 \pm 1.4$ \\
PA\tablenotemark{a} [deg] & $90.7 \pm 0.1$ & $90.5 \pm 0.6$ \\
$v_r$\tablenotemark{b}\,$[\kms]$ & $+13.651 \pm 0.003$ & $+13.649 \pm 0.015$ \\
$\mu_\alpha\,[''~{\rm yr}^{-1}]$  & $-0.004 \pm 0.002$ & $-0.028 \pm 0.010$ \\
$\mu_\delta\,[''~{\rm yr}^{-1}]$ & $-0.044 \pm 0.002$ & $-0.051 \pm 0.008$ \\
\enddata
\tablenotetext{a}{For comparison with the stellar orbits, we note that the position angle of the ascending node $\Omega_\mathrm{disk}$ is $90^\circ$ offset from our PA convention, i.e. $\Omega_\mathrm{disk} = \mathrm{PA} + 90^\circ \approx 180.6^\circ$.}
\tablenotetext{b}{LSRK reference frame. In the barycentric reference frame, the disk systemic velocity is $+28.34\,\kms$.}
\tablecomments{The 1D marginal posteriors are well-described by a Gaussian, so we report symmetric error bars here.}
\end{deluxetable}

Compared to previous similar work, the modeling of \obj\ is considerably more computationally expensive. This is primarily a consequence of the large physical size of the disk, which makes the ray-tracing step substantially more time-consuming. The inference for an individual spectral line takes $\sim$10,000 CPU hours parallelized across 26 cores on the Harvard Odyssey cluster.  Given that expense, and the fact that the $^{12}$CO line is clearly contaminated by local cloud material, we restrict our analysis to {\it independent} inferences of the model parameters based on the $^{13}$CO and C$^{18}$O datasets. For expediency, we only model the data averaged to 25 channels of $0.4\,\kms$ width. Experiments modeling a subset of the channels at higher resolution (e.g., using every third 0.17\,\kms-wide channel) yielded similar results.

\begin{figure*}[ht!]
\begin{center}
  \includegraphics{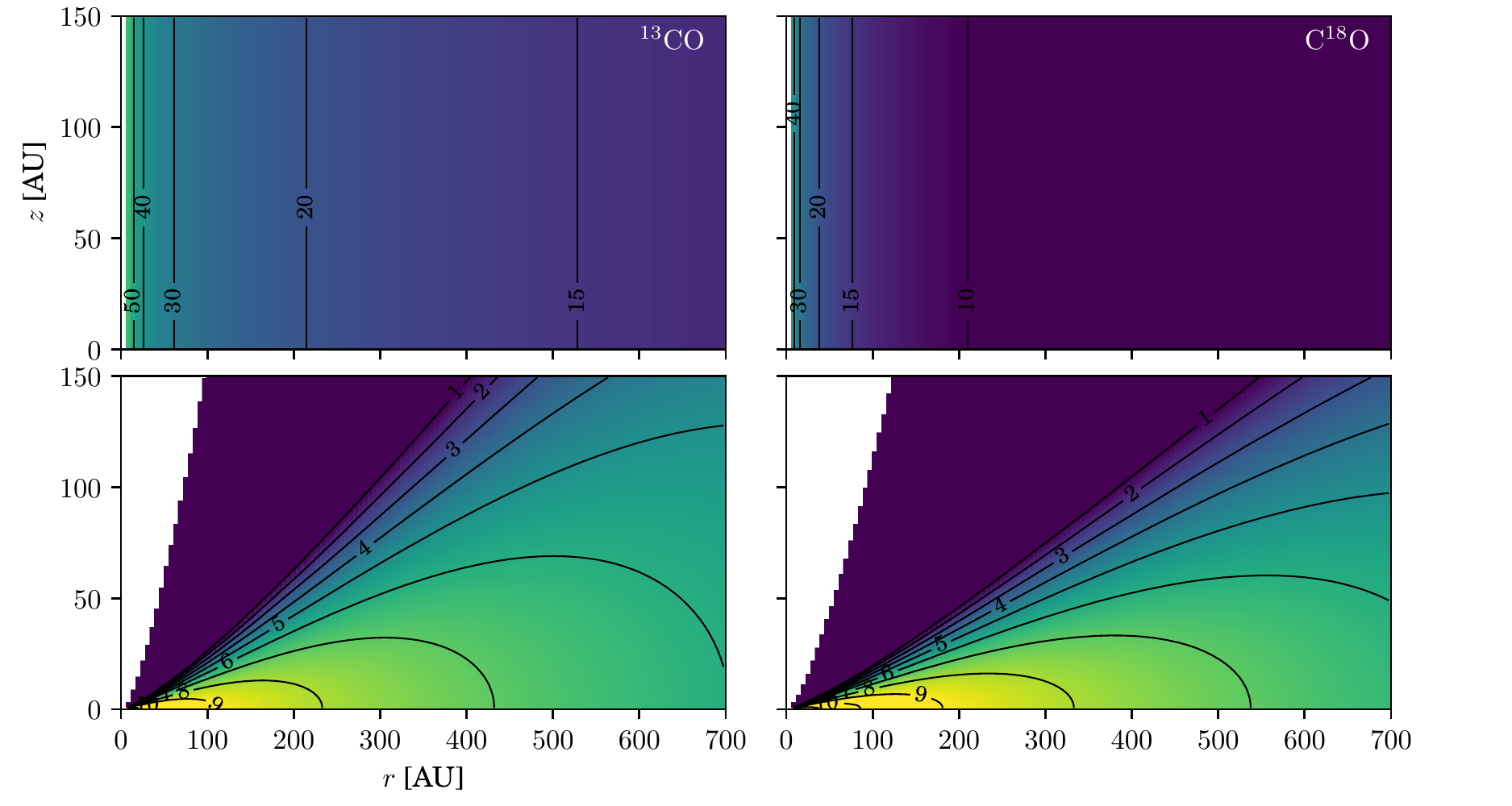}
  \figcaption{The maximum likelihood 2D temperature (top) and density (bottom) disk structures inferred using the \thirteen\ (left) and \eighteen\ (right) transitions. The temperature contours are in units of K. The density plots show the total gas density ($\rho_\mathrm{gas}$) and are in units $\log_{10} \mathrm{cm}^{-3}$. The color scales are normalized to the same limits for both transitions.
  \label{fig:temp_dens}}
  \end{center}
\end{figure*}

\begin{figure}[ht!]
\begin{center}
	\includegraphics{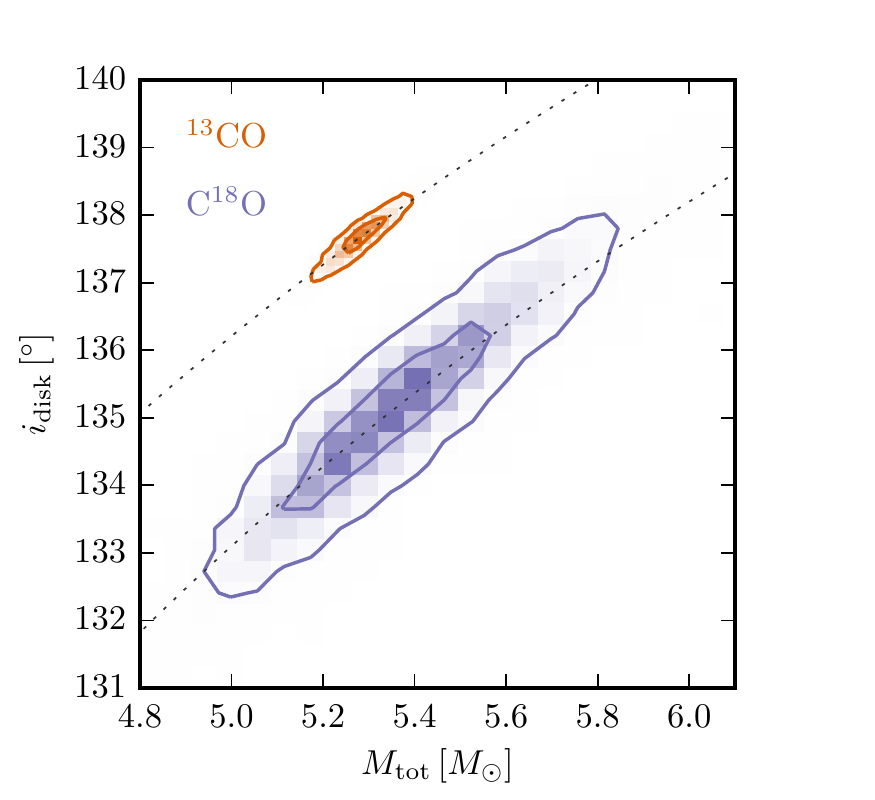}
  \figcaption{Posterior distributions for the model parameters fit to the \thirteen\ and \eighteen\ data independently, showing 1 and 2 $\sigma$ contours. Dashed lines indicate constant values of $M_\mathrm{tot} \sin^2 i_\mathrm{disk}$. \label{fig:posterior}}
  \end{center}
\end{figure}

The parameter values inferred from each spectral line dataset are summarized together in Table~\ref{table:components}.  A comparison of the data and the best-fit models (and associated residuals) is presented in the form of channel maps in Figures~\ref{fig:13co} and \ref{fig:c18o} for $^{13}$CO and C$^{18}$O, respectively.  While overall the models successfully reproduce the observed emission, there are some interesting residuals, namely, an excess of emission in the center of the disk for the channels between 13.1--14.3\,\kms, seen in both \thirteen\ and \eighteen. We will return to a discussion of a potential origin of those residuals in Section~\ref{sec:eclipses}.

Motivated by the presence of the aforementioned residuals, we explored more sophisticated disk models, including a model with a vertical temperature gradient and CO depletion due to freeze-out and photodissociation \citep[after][]{rosenfeld13a}, as well as a flexible temperature model parameterized to mimic more sophisticated (and computationally expensive) protoplanetary disk models \citep{kamp04,jonkheid04}. However, we found that neither of these models resulted in a more satisfactory fit to the data as measured by visual inspection and the Akaike information criterion \citep[AIC;][]{akaike73}. Encouragingly, however, they still yielded similar estimates of $M_\mathrm{tot}$ as the standard model, which gives us confidence that the disk-based dynamical mass is sufficiently robust to choice of parameterization for the temperature and density structures.

While the inferred physical structures inferred from each line are in mild disagreement, as might be expected for this simple parameterization, the spatial values of the inferred temperature and density in the disk are for the most part quite similar. To illustrate this, we plot the 2D temperature and density profiles inferred from each transition in Figure~\ref{fig:temp_dens}. We attribute the small differences in the structure parameters to the different layers of the disk probed by the \thirteen\ and \eighteen\ transitions: the more optically thick \thirteen\ probes the upper layers of the disk atmosphere while the more optically thin \eighteen\ transition is more sensitive to the colder, denser, layers of the disk midplane. In Figure~\ref{fig:posterior}, we plot the marginalized posteriors for both transitions in the \{$M_\mathrm{tot}$, $i_\mathrm{disk}$\}-plane. Interestingly, the different transitions deliver different inclinations ($\Delta i_\mathrm{disk} = 2.5 \pm 1.4^\circ$), which we attribute to the previously mentioned model deficiencies and the fact that the \thirteen\ and \eighteen\ transitions probe different layers in the disk. With more computational power, it would be worthwhile to explore a joint fit to both transitions to see if a single, more sophisticated, disk structure could adequately fit both transitions simultaneously.

Nevertheless, both transitions yield consistent constraints on the total stellar mass, which is the most relevant parameter to our stated goals. The robustness of the dynamical mass technique is primarily because the kinematic morphology of the line emission (i.e., the distribution of the emission in position--velocity space) is not strongly dependent on the temperature and density structure of the disk, but is a rather strong function of $M_\mathrm{tot}$ and $i_\mathrm{disk}$. When the disk is spatially resolved, the dependence of $M_\mathrm{tot}$ on $i_\mathrm{disk}$ is considerably diminished. 

We combine the inferred total masses from \thirteen\ and \eighteen, weighted by their uncertainties, to find $M_\mathrm{tot} = 5.29 \pm 0.06\,M_\odot$. The uncertainty in the distance to \obj\ \citep[$388 \pm 5\,$pc;][]{kounkel17} linearly translates into a mass uncertainty, and so we convolve an additional 1.3\% mass uncertainty with this posterior to arrive at $M_\mathrm{tot} = 5.29 \pm 0.09\,M_\odot$, which we report as the total mass estimate. Because the inferred disk inclinations are mutually inconsistent, we adopt a weighted average for the mean inclination and assume a large systematic uncertainty, resulting in a final estimate of $i_\mathrm{disk} = 137.6 \pm 2.0^\circ$. Our CO results are broadly consistent with that determined by \citet{fang17}, who measure the disk inclination to be $\sim$35--40$^\circ$ (modulo the absolute inclination of the disk).

\begin{figure*}
\begin{center}
  \includegraphics{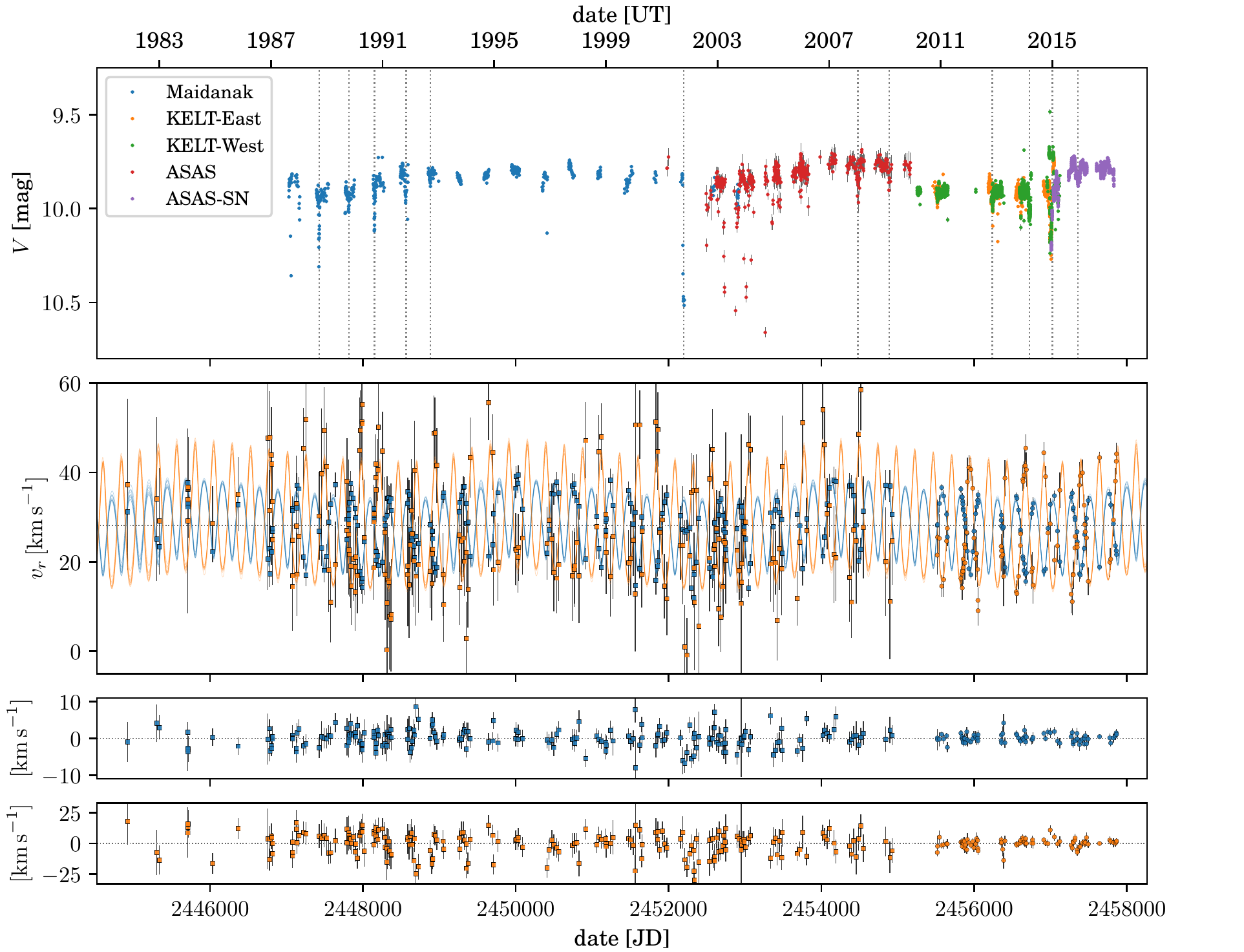}
  \figcaption{\emph{top}: photometric observations of GW~Ori from 1987 until mid 2017. All photometric observations displayed here are in the $V$-band (ASAS, ASAS-SN, and Maidanak) or a broader filter (KELT) which has been shifted to align with $V$-band where the time series overlap. \emph{bottom}: primary (blue) and secondary (orange) radial velocities overlaid with several realizations of the most probable orbits, to show uncertainty in the orbit. Reticon velocities are shown with squares, TRES velocities are shown with circles, and the dotted line represents the center-of-mass velocity. Residuals for this orbit are shown in the panels below.
  \label{fig:orbitboth}}
\end{center}
\end{figure*}

\begin{figure}
\begin{center}
  \includegraphics{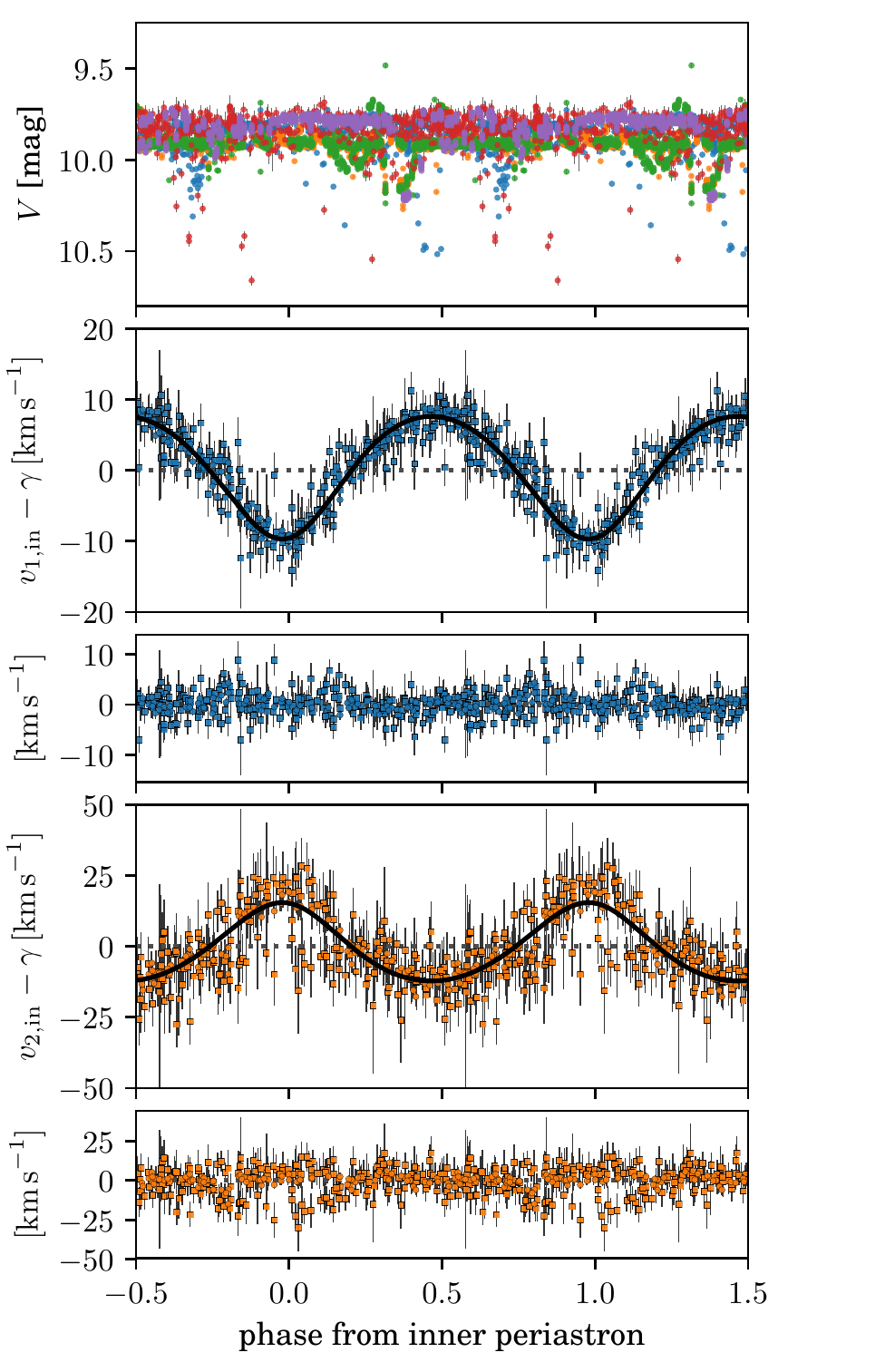}
  \figcaption{\emph{top}: $V$-band light curve phased to the inner orbital period. No discernible correlation is detected. The colors are the same as in the top panel of Figure~\ref{fig:orbitboth}. \emph{bottom}: RV measurements of \obj\ and best-fit model for the inner orbit, after subtracting the motion due to the outer orbit.  \label{fig:orbitin}}
\end{center}
\end{figure}

\begin{figure}
\begin{center}
  \includegraphics{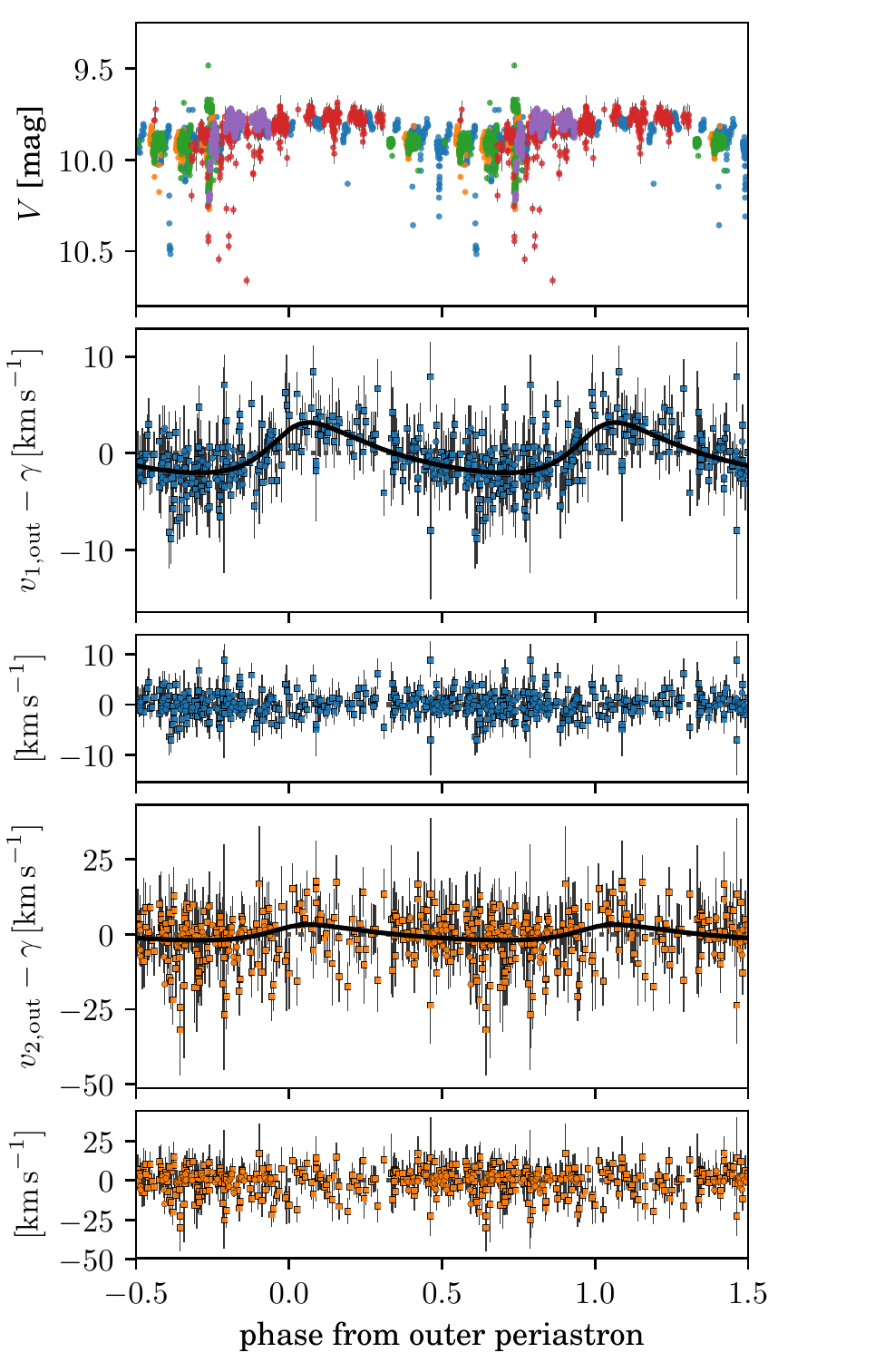}
  \figcaption{\emph{top}: $V$-band light curve phased to the outer orbital period, showing that the mean flux level oscillates by 0.2 mag over the course of the outer orbit, and is lowest during apoastron (phase 0.5). The colors are the same as in the top panel of Figure~\ref{fig:orbitboth}. \emph{bottom}: RV measurements of \obj\ and best-fit model for the outer orbit, after subtracting the motion due to the inner orbit.
  \label{fig:orbitout}}
\end{center}
\end{figure}

\subsection{An Updated Model of the Stellar Orbits \label{sec:orbit}}

In this section, we present an orbital fit to the RVs determined in \S\ref{subsec:spectroscopy}, and then explore a joint fit to the RV data and the astrometric measurements of \citet{berger11}.  In both cases we fit a hierarchical triple model and solve for the elements of the inner and outer orbits simultaneously, assuming the inner binary acts as a point mass in the outer orbit. To address possible systematic offsets in the RV datasets, we derive three offset terms: (1) $\Delta v^{\rm TRES}$, applied as a shift to all TRES RVs to place them on the DS reference frame; (2) $\Delta v_2^{\rm DS}$, to allow for an offset between the primary and secondary DS velocities, possibly caused by a mismatch between the template parameters and those of the true stars; and (3) $\Delta v_2^{\rm TRES}$, a similar primary/secondary offset for TRES. The residuals from our initial fit indicated that our formal velocity uncertainties are underestimated, and so the uncertainties on each measurement are scaled to achieve a reduced $\chi_\nu^2 = 1$ for our final solution.

The period of the inner orbit is consistent with that of \citet{mathieu91} and \citet{fang14}; however, due to the SB2 nature of the system, most other orbital parameters are significantly different. We find a larger semi-amplitude for the primary, $K_{\rm A} = 8.36 \pm 0.14~\kms$, a mass ratio of $q \equiv M_\mathrm{B} / M_\mathrm{A} = 0.60 \pm 0.02$, and a statistically significant eccentricity $e_\mathrm{in} = 0.13 \pm 0.02$. The outer orbit has an orbital period of \mbox{$P_\mathrm{out} = 4218 \pm 60$ days} (\mbox{11.5 years}), and a significant eccentricity, $e_\mathrm{out} = 0.22 \pm 0.09$. We find offset terms statistically inconsistent with zero: a small but non-negligible offset between the DS and TRES zeropoints of $0.49\;\kms$ and larger offsets for the secondary velocities of $8.77\;\kms$ and $6.41\;\kms$, for the DS and TRES RVs, respectively. Given the large intrinsic linewidth ($v \sin i \approx 40\;\kms$), these large offsets can reasonably be ascribed to template mismatch. The systemic velocity inferred from the RV fit is nicely consistent with the systemic velocity of the circumtriple disk from the ALMA data. All parameters of the RV fit are listed in the first column of Table~\ref{tab:elements}.  The full orbit as a function of time is shown in the second panel of Figure~\ref{fig:orbitboth}. Graphical representations of our observations and the inner and outer orbit models as a function of orbital phase are shown in Figure~\ref{fig:orbitin} and Figure~\ref{fig:orbitout}, respectively.

Although there are only three epochs of published astrometry in \citet{berger11}, these points may still help constrain the parameter space of possible orbits. Therefore, we explore a joint RV-astrometric analysis built upon a model of the ``three-dimensional orbit'' following \citet{murray10}, which adds new model parameters including the semi-major axis, orbital inclination, and position angle of the ascending node for both the inner and outer orbits. For a likelihood function, we combine the $\chi^2$ RV likelihood and a new $\chi^2$ likelihood for the angular separation and position angle measurements of the B and C components relative to A. As with the disk analysis, we also use a geometric prior on the orbital inclinations. For their last measurement epoch (2005), \citet{berger11} report an alternate position for the C component, and so we also perform a separate fit for this scenario.

\begin{figure*}[ht!]
\begin{center}
  \includegraphics{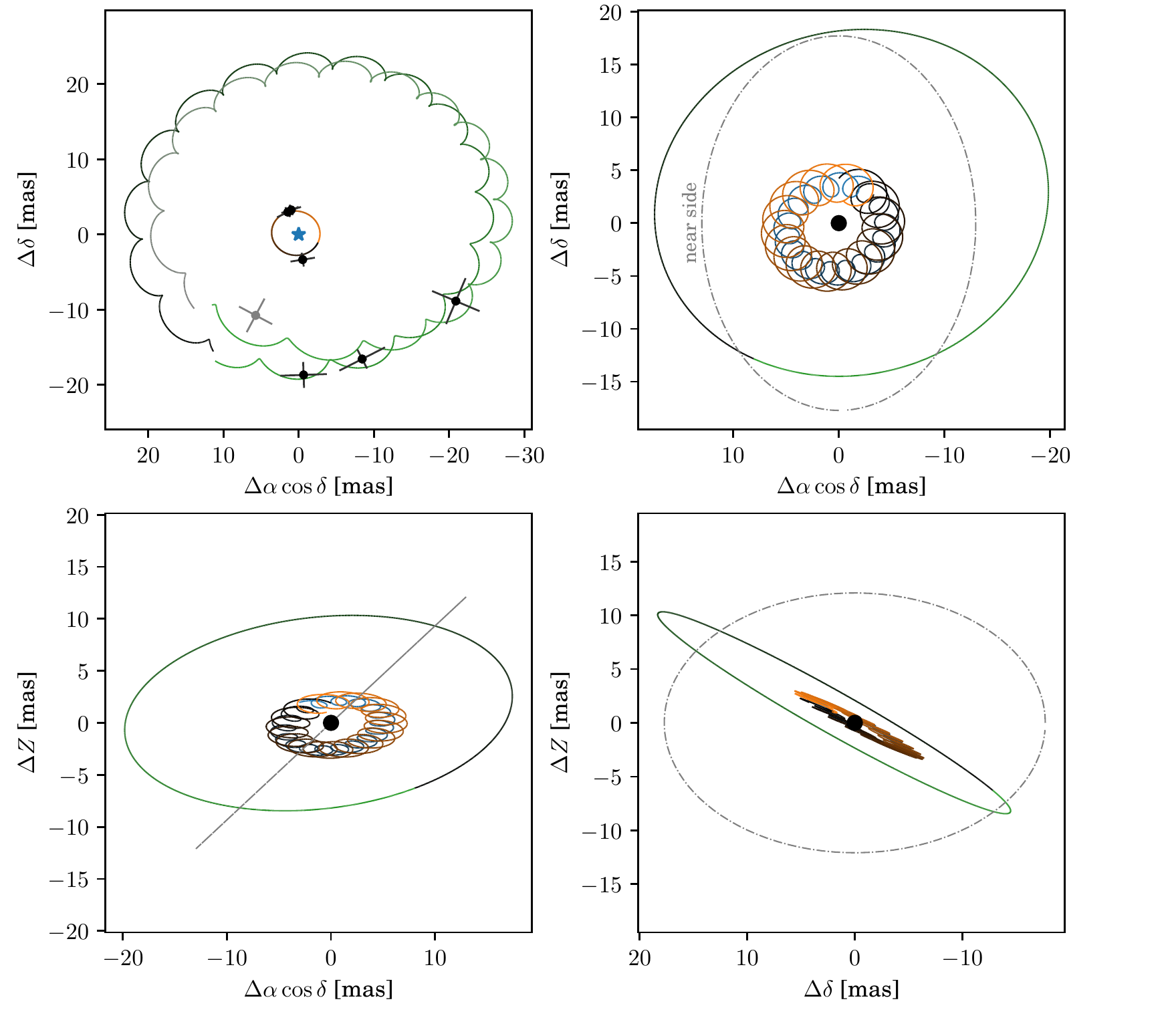}
  \figcaption{Orbits from the joint RV-astrometric fit shaded according to their phase, where black represents periastron and color hue increases with orbital phase. \emph{top left}: orbits relative to the primary star A, in the plane of the sky, with the three epochs of astrometry from \citet{berger11}. The light grey data point and outer orbit represent the fit to the ``alternate'' position for C. The following three plots are relative to the center of mass of the system. A fictitious particle at 7~au on a circular orbit coplanar with the circumtriple disk is shown as grey dashed line. \emph{top right}: the sky plane. For future discussion in \S\ref{sec:eclipses}, we label the side of the disk nearest to the observer. \emph{bottom left}: looking down the North axis. \emph{bottom right}: looking down the East axis. Positive Z points towards observer.
  \label{fig:astro}}
\end{center}
\end{figure*}

The jointly-constrained parameters are in the second and third columns of Table~\ref{tab:elements}.  A graphical representation of the orbit is shown in Figure~\ref{fig:astro}. With the addition of the astrometric dataset, we can measure the individual stellar masses and the inclinations of the orbital planes, which are also listed in Table~\ref{tab:elements}. Depending on whether the original or alternate position for C is used, we find the total stellar mass to be $M_\mathrm{tot} = 5.7 \pm 0.7\,M_\odot$ or $6.1 \pm 0.9\,M_\odot$, respectively. Both measurements are consistent with the $M_\mathrm{tot}$ independently measured with the disk-based analysis ($M_\mathrm{tot} = 5.29 \pm 0.09\,M_\odot$).

\begin{deluxetable*}{lccc}
\tablewidth{0pc}
\tablecaption{Orbital elements of \obj. \label{tab:elements}}
\tablehead{
\colhead{~~~~~~~~~~~~Parameter~~~~~~~~~~~~} & \colhead{RV} & \colhead{RV + astrometry} & \colhead{RV + astrometry\tablenotemark{$\dagger$}} 
}
\startdata
\multicolumn{4}{c}{Inner orbit} \\
\noalign{\vskip 1pt}
\noalign{\hrule}
\noalign{\vskip 3pt}
$P$ [days]\dotfill                        &    241.49~$\pm$~0.05\phn\phn  & 241.50~$\pm$~0.05 & 241.49~$\pm$~0.04\\
$K_\mathrm{A}$ [\kms]\dotfill                 &        8.36~$\pm$~0.14 & 8.34~$\pm$~0.15 & 8.36~$\pm$~0.15\\
$q$ \dotfill & 0.60~$\pm$~0.02 & 0.60~$\pm$~0.02 & 0.60~$\pm$~0.02  \\
$a$ [au] \dotfill & \nodata & 1.25~$\pm$~0.05 & 1.27~$\pm$0.05 \\
$e$\dotfill                               &      0.13~$\pm$~0.02 & 0.13~$\pm$~0.01 & 0.13~$\pm$~0.01 \\
$i$ [deg] \dotfill & \nodata & 157~$\pm$~1 & 157~$\pm$~1 \\
$\omega_\mathrm{A}$ [deg]\dotfill                    &        196~$\pm$~7 & 197~$\pm$~7 & 196~$\pm$~6 \\
$\Omega$\tablenotemark{b} [deg]\dotfill                    &  \nodata      &  263~$\pm$~13 & 264~$\pm$~13 \\
$T_{\rm peri}$ [HJD$-$2,400,000]\dotfill   &      56681~$\pm$~4\phm{222} & 56682~$\pm$~4 & 56681~$\pm$~4 \\
$\gamma$ [\kms]\dotfill                   &    +28.31~$\pm$~0.19\phn\phs & +28.33~$\pm$~0.18 & +28.29~$\pm$~0.19\\
$\Delta v$ TRES\tablenotemark{a} [\kms]\dotfill & 0.49~$\pm$~0.24 & 0.52~$\pm$~0.23 & 0.47~$\pm$~0.23\\
$\Delta v_2$ Reticon\tablenotemark{a} [\kms]\dotfill & 8.77~$\pm$~0.65 & 8.75~$\pm$~0.67 & 8.73~$\pm$~0.66\\
$\Delta v_2$ TRES\tablenotemark{a} [\kms]\dotfill & 6.41~$\pm$~0.37 & 6.36~$\pm$~0.35 & 6.39~$\pm$~0.39\\
$M_\mathrm{A}$ [$M_{\odot}$] \dotfill & \nodata & $2.80_{-0.31}^{+0.36}$ & $2.94_{-0.40}^{+0.40}$\\
$M_\mathrm{B}$ [$M_{\odot}$] \dotfill & \nodata & $1.68_{-0.18}^{+0.21}$ & $1.77_{-0.24}^{+0.24}$\\
\noalign{\vskip 2pt}
\noalign{\hrule}
\noalign{\vskip 2pt}
\multicolumn{4}{c}{Outer orbit} \\
\noalign{\vskip 1pt}
\noalign{\hrule}
\noalign{\vskip 3pt}
$P$ [days]\dotfill                        &       4218~$\pm$~60\phn\phn & 4246~$\pm$~66 & 4203~$\pm$~60 \\
$K_{\rm AB}$ [\kms]\dotfill                &       2.47~$\pm$~0.25 & 2.38~$\pm$~0.23 & 2.50~$\pm$~0.24 \\
$a$ [au] \dotfill & \nodata & 9.19~$\pm$~0.32 & 9.15~$\pm$~0.35 \\
$e$\dotfill                               &      0.22~$\pm$~0.09 & 0.13~$\pm$~0.07 & 0.25~$\pm$~0.08 \\
$i$ [deg] \dotfill & \nodata & 150~$\pm$~7 & 144~$\pm$~9 \\
$\omega_\mathrm{AB}$ [deg]\dotfill                    &        307~$\pm$~18\phn & 310~$\pm$~21 & 310~$\pm$~12 \\
$\Omega$\tablenotemark{b} [deg]\dotfill                    &    \nodata    & 282~$\pm$~9 & 263~$\pm$~10 \\
$T_{\rm peri}$ [HJD$-$2,400,000]\dotfill   &      53560~$\pm$~565\phn\phn & 53911~$\pm$~260 & 53878~$\pm$~130 \\
$M_\mathrm{C}$ [$M_{\odot}$] \dotfill & \nodata & $1.15_{-0.23}^{+0.40}$ & $0.99_{-0.18}^{+0.35}$\\
\noalign{\vskip 2pt}
\noalign{\hrule}
\noalign{\vskip 2pt}
\multicolumn{4}{c}{Derived properties} \\
\noalign{\vskip 1pt}
\noalign{\hrule}
\noalign{\vskip 3pt}
Inner time interval [cycles]\dotfill            &             53.6 & \nodata & \nodata \\
Outer time interval [cycles]\dotfill            &             3.1 & \nodata & \nodata \\
$M_{\rm A} \sin i \;$[$M_{\sun}$]\dotfill            &     0.30~$\pm$~0.02 & \nodata & \nodata \\
$M_{\rm B} \sin i \;$[$M_{\sun}$]\dotfill            &     0.18~$\pm$~0.01 & \nodata & \nodata \\
$M_{\rm C} \sin i / (M_\mathrm{tot} / M_\odot)^{2/3} \; $[$M_{\sun}$]\dotfill            &     0.22~$\pm$~0.02 & \nodata  & \nodata \\
$M_\mathrm{tot}\;$ [$M_{\sun}$]\dotfill & \nodata & $5.7 \pm 0.7$ & $6.1 \pm 0.9 $ \\
\enddata
\tablenotetext{a}{We include parameters for a potential velocity offset between the primary and secondary radial velocities for each instrument. In principle, this term should be consistent with $0$; the non-zero value likely indicates that there is some moderate template mismatch between the secondary stellar spectrum and the synthetic spectrum used as a cross correlation template.}
\tablenotetext{b}{We follow the convention of the visual binary field and define the ascending node as the point where the secondary component (e.g., star B for the inner orbit, and star C for the outer orbit) crosses the plane of the sky moving \emph{away} from the observer.}
\tablenotetext{\dagger}{Fit using the ``alternate'' C position for the 2005 epoch of astrometry in \citet{berger11}.}
\end{deluxetable*}

To measure the degree of misalignment between the orbital planes and the circumtriple disk, we calculate the angle $\Phi$ between the angular momentum vectors of each orbit according to \citet{fekel81}
\begin{equation}
  \cos \Phi = \cos i_1 \cos i_2 + \sin i_1 \sin i_2 \cos(\Omega_1 - \Omega_2).
\end{equation}
We find that the mutual inclination between the disk and the inner orbit is $\Phi_\mathrm{in} = 44\pm5^\circ$ and the mutual inclination between the disk and the outer orbit is $\Phi_\mathrm{out} = 54\pm 7^\circ$; these values are similar if one uses the ``alternate'' C position ($\Phi_\mathrm{in} = 45 \pm 5^\circ$, $\Phi_\mathrm{out} = 50 \pm 6^\circ$). Such a large misalignment is surprising given the naive expectation that the stellar orbits and disk would be roughly co-planar.  Since these results only rest upon three astrometric epochs, however, there is a possibility that the large inferred mutual inclinations may be the result of unaccounted for systematic effects. In the next section, we use only the newly derived RVs and disk-based dynamical mass to formulate a more conservative estimate of the mutual inclinations. We advocate continued astrometric monitoring of the \obj\ system to further improve the three-dimensional orbit and definitively confirm the inclinations of the stellar orbits.

\subsection{Joint RV + Disk Constraints on Individual Component Masses \label{sec:joint}}

In this section, we combine the RV analysis with the disk-based constraints on the total stellar mass to independently infer the individual stellar masses of the \obj\ system without referencing the \citet{berger11} astrometry. We construct a joint likelihood function with the following five parameters: $M_\mathrm{A}$, $M_\mathrm{B}$, $M_\mathrm{C}$, $i_\mathrm{in}$, and $i_\mathrm{out}$. The RV constraints are sufficiently captured by the summary statistics $M_\mathrm{A} \sin^3 i_\mathrm{in}$, $M_\mathrm{B} \sin^3 i_\mathrm{in}$, $M_\mathrm{C} \sin i_\mathrm{out} / (M_\mathrm{tot} / M_\odot)^{2/3}$ and the covariances between them, which are well represented by a multivariate Gaussian distribution.\footnote{Note that we do not use additional constraints on $q_\mathrm{in}$ or other derived parameters, as this would amount to double-counting the RV constraints.} The disk constraint on the total stellar mass $M_\mathrm{tot}$ is well-represented by a Gaussian, as well. We enforce flat priors on the stellar masses and geometrical priors on the inclinations. We use the ensemble sampler MCMC \citep{goodman10,foreman-mackey13} with 20 walkers to explore the posterior for 50,000 iterations, burn 25,000 iterations, and assess convergence by ensuring the Gelman-Rubin statistic \citep{gelman14} is $\hat{R} < 1.1$ for all parameters.

\begin{deluxetable}{lcc}
\tablecaption{Joint constraints on stellar masses and orbital inclinations \label{table:component_masses}}
\tablehead{\colhead{Parameter} & \colhead{RV + astrometry} & \colhead{RV + disk}}
\startdata
$M_\mathrm{A}$ [$M_\odot$] & $2.80_{-0.31}^{+0.36}$ & $2.74_{-0.52}^{+0.15}$  \\
$M_\mathrm{B}$ [$M_\odot$] & $1.68_{-0.18}^{+0.21}$ & $1.65_{-0.31}^{+0.10}$ \\
$M_\mathrm{C}$ [$M_\odot$] & $1.15_{-0.23}^{+0.40}$ & $0.88_{-0.19}^{+0.85}$ \\
$i_\mathrm{in}$ [deg] & $157_{-1}^{+1}$ & $151_{-2}^{+1}$ \\
$i_\mathrm{out}$ [deg] & $150_{-7}^{+7}$ & $130_{-27}^{+28}$ \\
\enddata
\tablecomments{The RV + astrometry values are replicated from Table~\ref{tab:elements} for comparison purposes. We note that we are not able to infer the absolute inclination of the stellar orbits directly from the radial velocity data, so there are in fact alternate solutions for the RV + disk results that yield $i_\mathrm{alt} = 180^\circ - i$. These solutions would be inconsistent with the astrometric motion, however, so we opt to only report the solutions with $i \geq 90^\circ$.}
\end{deluxetable}

\begin{figure}[ht!]
\begin{center}
  \includegraphics{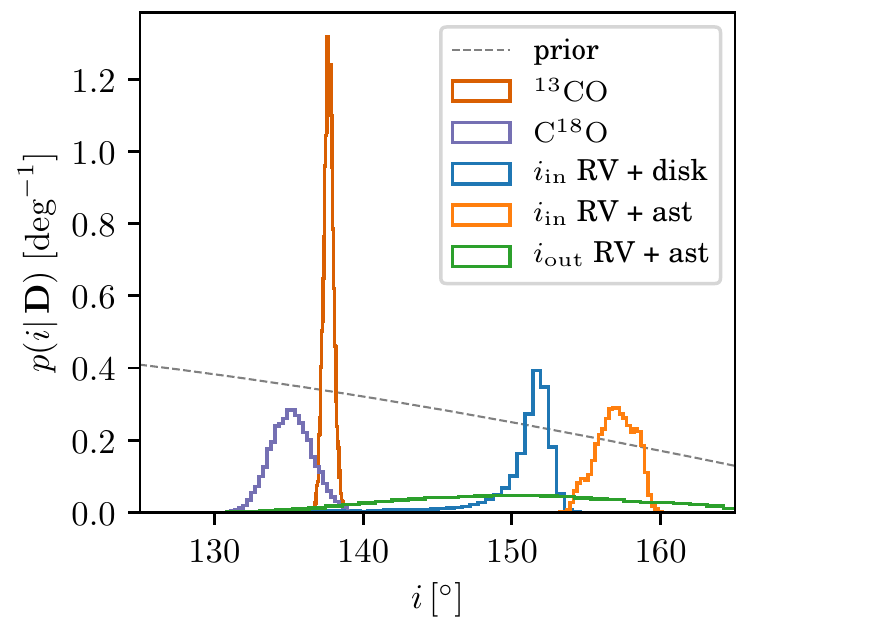}
  \figcaption{The inclination posteriors on the disk inclination, inner stellar orbit, and outer stellar orbit, as determined from various joint fits. Because $i_\mathrm{out}$ is essentially unconstrained by the RV + disk analysis, it is not plotted for aesthetic reasons. The geometric prior on inclination (uniform orientation of orbits in 3D space) is shown as a thin grey dotted line.
  \label{fig:incl_posterior}}
  \end{center}
\end{figure}

This analysis produces consistent but less precise constraints on the stellar masses as the joint RV + astrometric fits (see Table~\ref{table:component_masses}). Like the RV + astrometric analysis, the disk + RV analysis also indicates that the inner stellar orbit must be significantly misaligned with the disk, although the true misalignment is unknown: the difference between $i_\mathrm{in}$ and $i_\mathrm{disk}$ only provides a lower limit on the mutual inclination because the true mutual inclination must consider the position angles of the orbits as well. To highlight these findings, we overplot our newly derived constraints on the disk inclination and the orbits in Figure~\ref{fig:incl_posterior}. The measurements for the disk based on the \thirteen\ and \eighteen\ data indicate the lowest inclinations (nearest to edge-on, $i = 90\degr$). This is commensurate with the disk inclination measurements from \citet{fang17}, and so we consider these results to be robust. Interestingly, the constraints on $i_\mathrm{in}$ differ between the RV + astrometry and the RV + disk results at a significant level. We speculate that this difference might be due to unknown systematics in the astrometry or RV datasets, or potentially an error in our assumption of the distance to \obj. While the astrometry + RV analysis does not require a distance to the source, the disk-based analysis does require a distance in order to break the $M_\mathrm{tot} / d$ degeneracy. Although the exact degree of mutual inclination between the inner orbit and the disk is unknown, we conclude that it is at least 10\degr\ and potentially as high as 45\degr.

\subsection{Variability: Eclipses and Periodic Behavior \label{sec:eclipses}}

\begin{figure*}[!ht]
\includegraphics{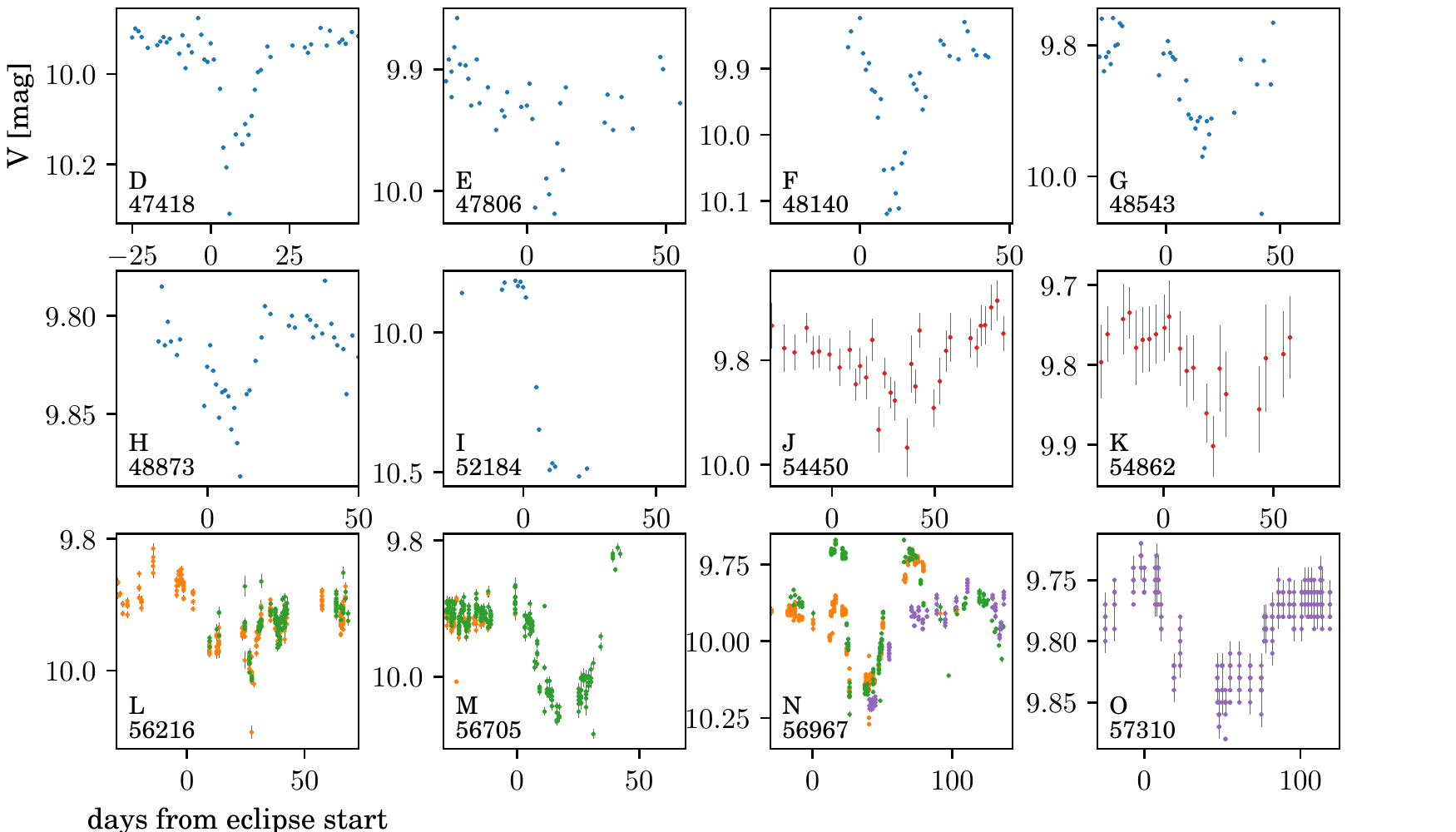}
\caption{A gallery of the eclipse events noted in Table~\ref{tab:eclipses}, labeled relative to the start of the eclipse. The colors are the same as in the top panel of Figure~\ref{fig:orbitboth}. Note that the y-axis scale is significantly different from panel to panel, with some eclipses as deep as 0.7 mag and others less than 0.1 mag.}
\label{fig:eclipse-gallery}
\end{figure*}

The first extensive lightcurve of \obj\ was published by \citet{shevchenko92}, based on several seasons of photoelectric photometry from Maidanak Observatory.  Among other modulations typical of young stars, those data revealed two deep ($\Delta V = 0.3$-0.4\,mag) eclipse-like events in 1988 and 1990 (Figure~\ref{fig:eclipse-gallery}, D and F). The span of time between these two events is exactly three orbital periods of the inner binary \citep[the only known orbit at the time;][]{mathieu91}. 
The correlation of the eclipses with the orbital phase led to the hypothesis that these features were Algol-like fadings, where the primary is obscured by material in the overflowing Roche lobe of the secondary. For that to work, the binary orbital plane must be viewed nearly edge-on ($i_\mathrm{in} \approx 90\degr$).  

Maidanak Observatory continued their extensive photometric monitoring of \obj\ until 1996 \citet{shevchenko98}. Those data identified two additional eclipses, one in 1991, and one in 1992 (Figure~\ref{fig:eclipse-gallery}, G and H), which occurred exactly three orbital periods after the 1990 event, but with a lower amplitude ($\Delta V = 0.1\,$mag). \citet{shevchenko98} noted that \obj\ appears to redden (in $V-R$ and $B-V$ colors) during all of these eclipses; however, insufficient precision was available to robustly constrain an associated extinction curve. Photoelectric observations by W.~Herbst reveal three additional eclipses of similar depth between 1982 and 1985 \citep[events A - C;][]{shevchenko98}. Those features also appear to be separated by integer multiples of the inner orbital period, lending further support to the Algol-like variable hypothesis. With their longer photometric time baseline, \citet{shevchenko98} also noted an overall decline of $\Delta V = 0.1$\,mag in the average brightness of \obj\ from 1986--1991.

\begin{deluxetable*}{clrrrrl}
\tablecaption{V-band Photometric Eclipse Catalog\label{tab:eclipses}}
\tablehead{\colhead{Label} & \colhead{UT Mid} & \colhead{Start} & \colhead{End} & \colhead{Duration} & \colhead{Depth} & \colhead{Telescope} \\
& & \colhead{JD} & \colhead{JD} & \colhead{days} & \colhead{mag} & }
\startdata
A & 1982 Oct & \nodata & \nodata & \nodata & \nodata & Herbst \\
B & 1983 May & \nodata & \nodata & \nodata & \nodata & Herbst \\
C & 1984 Nov & \nodata & \nodata & \nodata & \nodata & Herbst \\
D & 1988 Sep 13 & 2447418 & 2447435 & 17 & 0.40 & Maidanak \\
E & 1989 Oct 6 & 2447806 & 2447833 & 27 & 0.10 & Maidanak \\
F & 1990 Sep 5 & 2448140 & 2448161 & 21 & 0.35 & Maidanak \\
G & 1991 Oct 13 & 2448543 & 2448589 & 46 & 0.15 & Maidanak \\
H & 1992 Sep 7 & 2448873 & 2448893 & 20  & 0.08 & Maidanak \\
I & 2001 Oct 1 & 2452184 & $\geq$2452215 & $\geq$31 & 0.70 & Maidanak \\
J & 2007 Dec 15 & 2454450 & 2454508 & 58 & 0.10 & ASAS \\
K & 2009 Jan 30 & 2454862 & 2454912 & 50 & 0.15 & ASAS \\
L & 2012 Oct 15 & 2456216 & 2456259 & 43 & 0.11 & KELT \\
M & 2014 Feb 16 & 2456705 & 2456744 & 39 & 0.20 & KELT \\
N & 2014 Nov 5 & 2456967 & 2457080 & 30-130 & 0.25 & KELT/ASAS-SN \\
O & 2015 Oct 14 & 2457310 & 2457405 & 95 & 0.10 & ASAS-SN \\
\enddata
\end{deluxetable*}

Given the apparent connection of these eclipse events to the orbital architecture of \obj, we set out to conduct a more comprehensive analysis of photometric variability in this system. We combined the complete Maidanak photometry catalog (1987--2003) with data from the ASAS, KELT, and ASAS-SN surveys to construct a lightcurve spanning 1987 to 2017. This composite lightcurve was manually searched for new eclipse events.  Eight new eclipses were identified, bringing the total number to 15: they are listed in Table~\ref{tab:eclipses}, marked in Figure~\ref{fig:orbitboth}, and shown in greater detail in Figure~\ref{fig:eclipse-gallery}. Given the sometimes considerable noise in this lightcurve, we have only considered an event to be an eclipse if it shows multiple consecutive photometric points that deviate significantly from the running average (large dips that span only a single epoch are likely spurious). The identified eclipses are similar to the events identified by \citet{shevchenko92,shevchenko98}, ranging from 0.08 to 0.70\,mag in depth and lasting 10 to \mbox{50 days}. 

The longer time baseline for the combined photometric catalog indicates that the eclipse events are not exclusively periodic.  The 1988, 1990, and 1992 eclipses noted by \citet{shevchenko98} do indeed occur on integer multiples of the inner period of \mbox{241 days}. However, a reanalysis of the Maidanak photometry reveals an additional likely eclipse event in 1989 (E) which is three orbital periods apart from the 1991 event (G), but not at the same orbital phase as the D, F, and H events: rather, they are offset by about a third of an orbital period. Eclipses seen in the data from the more recent surveys show a similar behavior. For example, eclipses L and M are exactly two orbital periods apart, while eclipses N and O are about a quarter period early or late. When the full lightcurve is phased to the inner period, as in Figure~\ref{fig:orbitin}, the multitude of eclipse phases becomes readily apparent.

Aside from the eclipse events, the combined \obj\ lightcurve exhibits a striking sinusoidal variability (0.20\,mag peak-to-trough) that is clearly phased with the outer orbital period of 11.5\,yr (Figure~\ref{fig:orbitout}). We must be careful when evaluating this oscillation mode, since it stretches across data acquired from several different instruments. Moreover, the KELT dataset was not taken in $V$-band: it was shifted to match the overlapping ASAS-SN observations, meaning that it has a potentially problematic zeropoint uncertainty. With these caveats in mind, the clear rising and falling trends are seen within each individual dataset without the need for any vertical shifts, suggesting that this modulation is likely real. The earlier photoelectric observations of W.~Herbst, stretching back to 1983, also clearly phase up with the expected sinusoidal variation. This situates the long term dimming seen by \citet{shevchenko98} as part of an 11.5\,yr period brightness oscillation.

When considering the phase-folded lightcurve on the outer (AB--C) orbital period, it appears as if the deep eclipses preferentially occur between phases 0.4--0.8 from periastron. However, the three eclipses in the photoelectric observations of W.~Herbst fall closer to 0.0 phase. Taken together, this suggests that the apparent clustering of deep eclipses could be affected by the seasonal sampling in the dataset.

The 11.5\,yr variability reaches a minimum flux level near apoastron (phase = 0.5) of the outer orbit. Since the light from the A--B binary dominates the total optical flux from the system, it must be one or both of these stars that are either being partially occulted by circumstellar material. Due to the fact that the disk is more inclined than the stellar orbits, apoastron corresponds to the time when the A--B pair comes closest to being screened by the inner edge of the near side of the disk (see Figure~\ref{fig:astro}, top right panel). From dynamical arguments, we expect the inner edge of the disk to be truncated out to 2--3 times the semi-major axis of the tertiary \citep{artymowicz94}, which corresponds to $\sim$20--30\,au in radius. Even with the foreshortening from the relative inclinations, the inner edge of the disk would not occult the A--B pair at apoastron unless the inner edge of the disk were very puffed up with a vertical extent $\gtrsim 10\,$au. Given the gradual dimming, it may be more likely that the A--B pair is screened by tenuous material residing inside this truncation radius, such as micron-sized dust within the cleared region. There is circumstantial support for this interpretation from the variable infrared SED, which \citet{fang14} interpret as an indication of a variable reservoir of small grains near the A--B pairing which is cleared and replenished due to the actions of the tertiary.

We find some \thirteen\ and \eighteen\ emission that exceeds predictions from the most probable standard disk model at locations consistent with this near edge of the disk, but located at or near the systemic velocity (13.1--14.3\,$\kms$ LSRK). Those residuals could simply be an artifact from using an insufficiently complex disk model, but they may instead very well be probing the source of the eclipses, the long term dimming, or both. There is an outstanding question from this analysis as to what the disk looks like on the scales of the inner orbit. The ALMA observations do not have sufficient angular resolution to probe the disk at the physical scales corresponding to the tertiary orbit, so the distribution of solids within the \obj\ disk remains relatively unconstrained. With its longest baselines, ALMA would have the spatial resolution ($0\farcs02$) to probe the disk on 8\,au scales, more than sufficient to resolve a cleared region consistent with the tertiary orbit ($\gtrsim$18\,au in diameter).

Because the eclipse events are not synchronized with the outer orbital period, it is not immediately clear whether they share the same physical origin as the longer-term brightness variations. The new dynamical constraints derived earlier indicate that the A--B orbit is {\it not} edge-on, and so we must consider alternatives to the Algol mechanism. Any theory that seeks to explain these inner eclipses must account for several pieces of evidence, in addition to the updated orbital configuration. The eclipses span \mbox{10--50 days} in duration, are of variable depth (between 0.08 and 0.70 magnitudes), appear to be consistent with reddening by dust, are not strictly periodic with the inner period, and seem to occur at all phases of the outer orbital period. Moreover, the spectra (unwittingly) taken during times of eclipse show no obvious changes in spectral features beyond the normal variability described by \citet{fang14}. We speculate that these quasi-periodic eclipses may be due to an unstable circumbinary disk around A--B, or possibly the result of eclipses of A by accretion streams onto either the individual circumstellar disks of A or B.

\begin{figure}[!ht]
\includegraphics{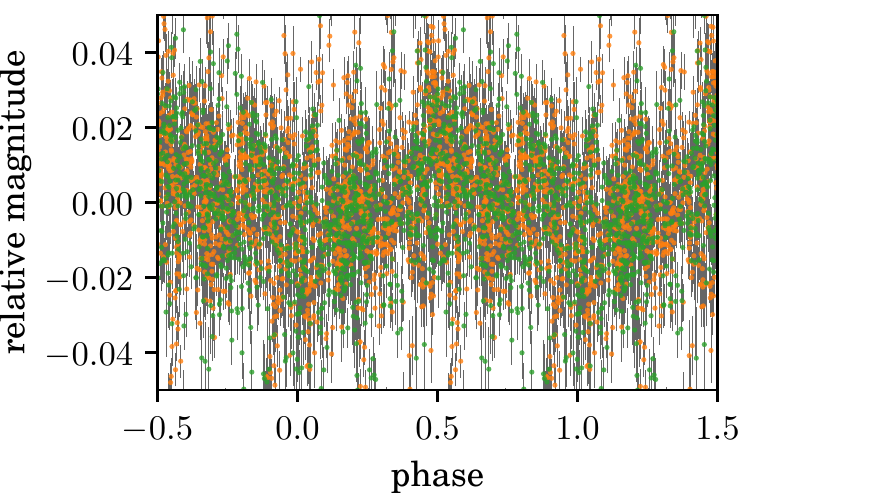}
\caption{The KELT photometric observations of \obj, with the three eclipses removed, phased to the 2.93 day period recovered from our LS analysis. Legend as in Figure~\ref{fig:orbitboth}}
\label{fig:phased}
\end{figure}

Finally, we searched for additional periodicty in the lightcurve beyond the inner and outer orbital periods. After excluding eclipses, we used the Lomb-Scargle (LS) periodicity search algorithm \citep{lomb76,scargle82}  \citep[within the VARTOOLS analysis package;][]{hartman16} to search for periodic modulations from \mbox{1.1 to 100 days} in the high-cadence KELT dataset. The most significant period we recover is \mbox{$2.93 \pm 0.05$\,days}, shown in Figure~\ref{fig:phased}. We suggest that this corresponds to the rotation period of the primary, although we note that \citet{bouvier90} and \citet{fang14} derived alternate rotation periods of \mbox{3.3 days} and \mbox{5.0--6.7 days}, respectively. Future high-precision, high-cadence observations of \obj\ will help to unambiguously identify the rotation period of A.

\section{Discussion} \label{sec:discussion}

With the newly derived component masses now established, we turn to discussion of the photospheric properties of the stars and the age of the \obj\ system. With these in place, we discuss the system architecture in the context of other multiple systems.

\subsection{Age and Photospheric Properties}

In order to place \obj~A and \obj~B on the HR diagram, we require updated measurements of their luminosities. To obtain those, we assembled an SED of \obj\ from the same sources listed in \citet{fang14}, i.e., the $U B V R_\mathrm{C} I_\mathrm{C}$ photometry from \citet{calvet04} and the $JHK_\mathrm{s}$ photometry from the 2MASS survey \citep{skrutskie06}. We then manually adjusted the 2MASS $J$ and $H$ fluxes down by 5\% and 10\%, respectively, to account for the approximate contamination in those bands from star C based on the flux ratios from \citet{berger11}. We fitted the SED using a two-component model based upon the NextGen atmospheres \citep{hauschildt99} and the \citet{cardelli89} reddening law with the following constraints: the distance to the source ($388\pm 5$\,pc), the spectroscopically-determined flux ratio at $5187$\,\AA\ ($0.25 \pm 0.05$), the primary $T_\mathrm{eff}$ ($5700\pm 200$\,K), and the $H$-band flux ratio \citep[$f_\mathrm{B} / f_\mathrm{A} = 0.57 \pm 0.05$;][]{berger11}. That analysis yields an extinction of $A_V = 1.2 \pm 0.2$\,mag, secondary $T_\mathrm{eff} = 4900 \pm 200$\,K, luminosities of $L_\mathrm{A} = 32.5 \pm 5.0\,L_\odot$ and $L_\mathrm{B} = 12.8 \pm 2.4\,L_\odot$, radii of $R_\mathrm{A} = 5.90 \pm 0.18\,R_\odot$ and $R_\mathrm{B} = 5.01 \pm 0.22\,R_\odot$, and a $V$-band flux ratio of $0.33\pm0.04$.

\begin{figure}[htb]
\begin{center}
  \includegraphics{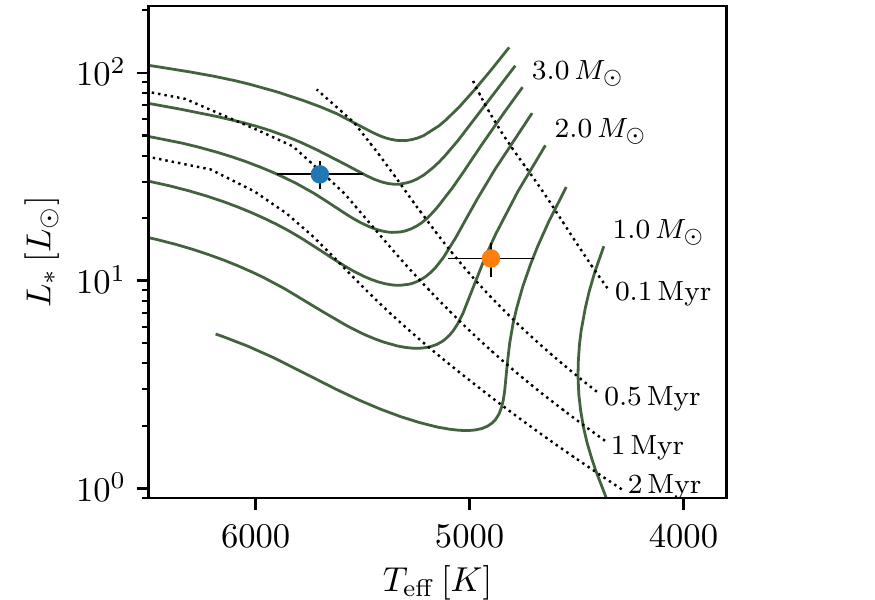}
  \figcaption{\obj~A and B placed on the pre-main sequence HR diagram, with evolutionary tracks from \citet{choi16}. Mass tracks are in increments of $0.5\,M_\odot$ from $1.0\,M_\odot$ to $4.0\,M_\odot$, and isochrones label 0.1, 0.5, 1, 2 Myr ages.
  \label{fig:PMS}}
  \end{center}
\end{figure}

Figure~\ref{fig:PMS} places the \obj\ A and B stars on the HR diagram, along with some representative stellar evolutionary tracks from the MIST models \citep{choi16}. The mass tracks start at $0.5\,M_\odot$ and increase in increments of $0.5\,M_\odot$; the isochrones are at ages of 0.1, 0.5, 1.0, 2.0, and 3.0 Myr. The positions of A and B in this diagram are consistent with their measured dynamical masses at ages of 0.3--1.3 Myr. However, the evolutionary tracks would imply that the A component is older than the B component by at least 0.3\,Myr. This age discrepancy is likely not real, but rather the result of innaccuracies in the photospheric properties, evolutionary models, or both. The revised mass for \obj~A is significantly lower than previous estimates in the literature \citep{berger11,fang14}, which, combined with the revised photospheric properties, means that the age of \obj\ is now slightly older than was previously implied. Even with its lower mass and luminosity, however, it is still true that \obj~A is a Herbig Ae/Be precursor \citep[as was also noted by][]{fang14}, and will evolve to a main sequence late-B/early-A star.

\begin{figure}[htb]
\begin{center}
  \includegraphics{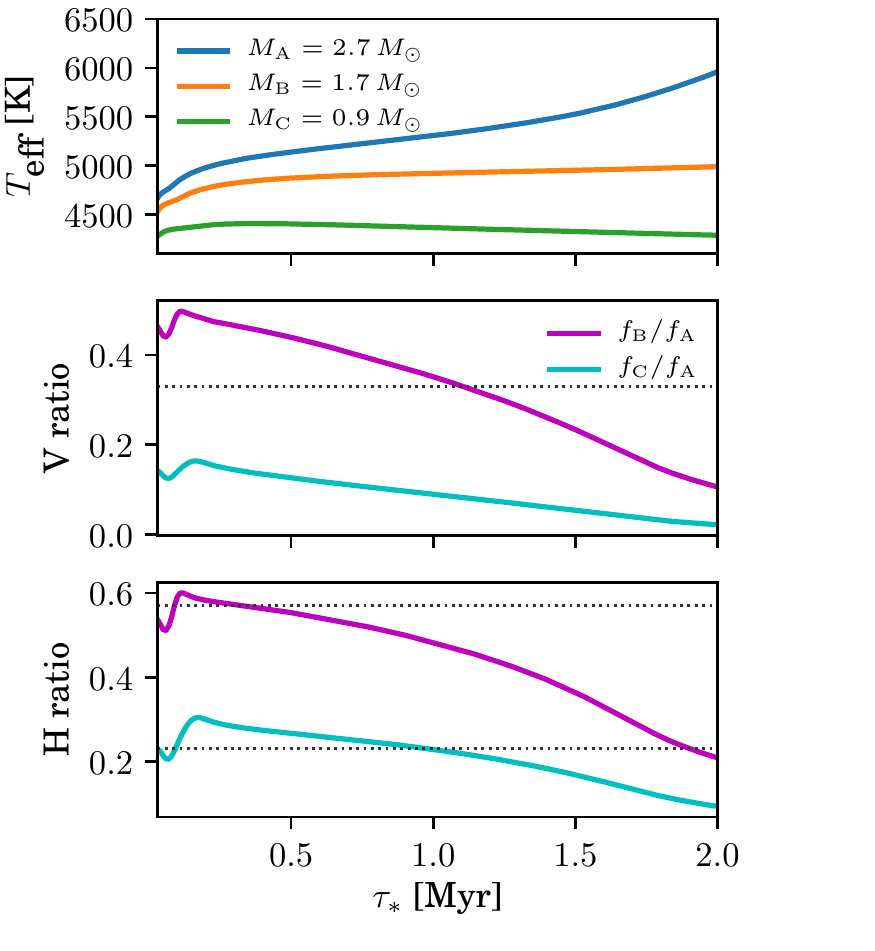}
  \figcaption{Relative photospheric properties of the \obj\ stellar components as a function of age, using the MIST pre-main sequence evolutionary models and assuming the stars are coeval. The measured effective temperatures for the primary and secondary ($T_\mathrm{eff} = 5700\pm200$\,K and $T_\mathrm{eff} = 4900\pm200$K, respectively), the $V$-band flux ratio ($f_\mathrm{B}/f_\mathrm{A} = 0.33\pm 0.04$), and the $H$-band flux ratios \citep[$f_\mathrm{B}/f_\mathrm{A} = 0.57 \pm 0.05$ and $f_\mathrm{C}/f_\mathrm{A} = 0.23 \pm 0.01$;][]{berger11} are shown here as dotted lines and are all roughly consistent with the model predictions for ages of 0.3 - 1.3 Myr.
  \label{fig:evolution}}
  \end{center}
\end{figure}

In Figure~\ref{fig:evolution}, we utilize the three MIST models nearest the best-fit masses for each of the stellar components to compute the evolution of the effective temperatures and the $V$- and $H$-band flux ratios, and compare these quantities to the existing photospheric measurements. Beyond the primary and secondary effective temperatures and $V$-band flux ratio reported in this work, there are two $H$-band flux ratio constraints from \citet{berger11},  $f_\mathrm{B}/f_\mathrm{A} = 0.57 \pm 0.05$ and $f_\mathrm{C}/f_\mathrm{A} = 0.23 \pm 0.01$ (computed as the weighted mean of all three of their epochs). In general, the measured effective temperatures and flux ratios agree well with the predictions from the models in the age range 0.3--1.3 Myr. The predicted $V$-band flux contribution for the C component is very small ($\lesssim 5\%$ of the total flux), explaining why we were unable to find optical spectroscopic signatures of this component even though it is nearly a solar mass star.

As noted by \citet{berger11}, it is possible that one, two, or all three of the stars might show excess $H$-band emission due to the presence of a circumstellar disk and/or accretion signatures above the photospheric emission of these stars. While we find that the measured photospheric properties are reasonably consistent without invoking such an excess, there is circumstantial evidence for the idea (beyond the eclipses). \citet{najita03} found blended CO fundamental emission peaks, which \citet{bast11} suggested might actually be the blended profiles of emission originating from physically distinct regions, e.g., individual circumstellar disks around the A and B stars. However, it may also be possible that the CO fundamental emission originates from the inner edge of the circumtriple disk, cleared by the $\sim$9\,au orbit of the tertiary (C).

Assuming that the 2.93 day period identified in the KELT photometry corresponds to the rotation period of the primary, we can use the inferred primary radius and the spectroscopic measurement of the rotational line broadening ($v \sin i_\mathrm{A}$) to infer its stellar obliquity, modulo its absolute orientation. Solving for the inclination and propagating the uncertainties in the respective parameters yields an obliquity of $i_\mathrm{A} = 23 \pm 3\degr$, which is in remarkable agreement (modulo the absolute orientation) with the inclination of the inner stellar orbit ($157 \pm 1\degr$) as determined from the RV + astrometry analysis (Sect.~\ref{sec:joint}).

\subsection{The \obj\ Triple System in Context}

The many unique datasets presented in this paper have enabled us to paint a detailed picture of the \obj\ system. It is young (0.3--1.3 Myr), contains a considerable amount of stellar mass ($M_\mathrm{tot} = 5.29\pm0.09\,M_\odot$), and hosts a massive disk ($M_\mathrm{disk} \approx 0.1\,M_\odot$), which makes it an extremely interesting system to study in the context of theories about star and planet formation, migration, and stability. 

We estimated the total disk mass of \obj\ using the results from our \thirteen\ and \eighteen\ modeling in \S\ref{sec:disk}, which we emphasize are very indirect measurements that rely upon uncertain conversion factors between \twelve\ and $\mathrm{H}_2$.
We find somewhat larger disk masses when modeling \eighteen\ compared to \thirteen\ ($0.095\,M_\odot$ vs. $0.020\,M_\odot$, respectively), which is in conflict with the finding of \citet{fang17} that \eighteen\ must be depleted relative to \thirteen. We attribute the differences in our disk masses to insufficiently complex models of disk structure and optical depth effects, and note that in general estimating disk masses from CO is notoriously difficult \citep{yu17}, although in our case it is encouraging that they are roughly consistent with estimates based on the dust continuum emission \citep[$0.1\,M_\sun$;][]{fang17}. In the context of the large disk mass survey by \citet{andrews13}, \obj's disk mass is slightly larger than the mean predicted value for its stellar mass, although still consistent with the large $1\sigma$ envelope in this relationship at high stellar masses. In light of this large disk mass, we investigate whether the disk is Toomre stable today. We use the more massive $M_\mathrm{disk}$ values from the \eighteen\ results to derive a lower bound on Toomre's Q parameter
\begin{equation}
Q(r) = \frac{c_s \Omega}{\pi G \Sigma} = \sqrt{\frac{k_\mathrm{B} M_\mathrm{tot}}{\pi^2 \mu m_\mathrm{H} G}} \sqrt{\frac{T(r)}{r^2 \Sigma^2(r)}}.
\end{equation}
For the range of disk parameters determined from our CO fitting, the minimum value is $Q \approx 100$ at $r \sim 300\,$au, which means that the disk is not currently undergoing a global gravitational instability ($Q \approx 1$).

Now we turn to a brief discussion of relevant analogues to the \obj\ system. There are now at least four circumbinary disks known around short period (10--20 days) eccentric binaries: UZ~Tau~E, V4046~Sgr, AK~Sco, and DQ~Tau \citep[][and references therein]{jensen07,rosenfeld12b,czekala15b,czekala16}. All four systems have their binary orbital plane and associated circumbinary disk aligned to within 3\degr. These findings agree well with the low mutual inclinations found for \emph{Kepler} circumbinary planets \citep{winn15}, and therefore have exciting implications for a large circumbinary planet occurrence rate \citep{li16}. Of course, severe selection effects are at work in both samples, and so care must be taken when extrapolating these results to the population at large. Within this context, \obj\ stands apart due to the fact that its stellar orbit is \emph{not} aligned with the disk. This may be a consequence of the larger stellar masses involved, the longer orbital period(s), and/or the existence of a close tertiary. 
Here we examine several other longer period systems that are characterized by their significantly non-zero mutual inclinations, and show that the large mutual inclinations found in \obj\ are not as unique when considering other longer period systems.

KH 15D is an eccentric ($e=0.6$) binary system with a slightly longer period than the aforementioned systems (\mbox{48 days}), and hosts a circumbinary dust ring misaligned by 10--20\degr\ from the stellar orbit \citep{chiang04,capelo12}. The eccentric stellar orbit and disk misalignment causes dramatic photometric eclipse events as stars are screened by the edge of the dust disk. The eclipses also come and go as the ring precesses about the binary. The edge of the occulting disk must be sharp, suggesting that the ring is confined to a narrow region by a planet at 4\,au. The disk and stellar orbital misalignment may be driven by dynamical interactions between the eccentric binary and the disk \citep[e.g.,][]{martin17,zanazzi18}.

Moving to still larger orbital separations, the frequency of disk-stellar orbital alignment becomes less clear, mainly due to incomplete orbital coverage. Consider GG~Tau~A, which is a triple system with circumstellar disks around each of three components, as well as a larger circumtriple disk. The circumtriple disk is composed of a dense ring containing 80\% of the mass and an outer gas disk extending up to 800\,au, and is of similar mass to that of the \obj\ disk \citep[$0.12\,M_\odot$;][]{guilloteau99}. The stellar architecture of GG~Tau~A is rather different than \obj, however. The primary star Aa has a mass of $0.78\,M_\odot$, and is situated in an ``outer'' orbit with another binary, Ab1--Ab2, which together have a combined mass of less than $0.7\,M_\odot$ \citep{dutrey16}. The orbital elements of the triple system still have some uncertainty. \citet{nelson16} make a dynamical argument that the outer orbit has a semi-major axis of 62\,au and is likely coplanar with the outer circumtriple ring; on the other hand, \citet{cazzoletti17} argue from disk dynamics that the disk and binary planes are misaligned by 20-$30^\circ$. Further astrometric observations are required to definitively characterize this system.

Recently, the transition disk system HD~142527 ($M_\ast = 2.0\,M_\odot$, $M_\mathrm{disk} \sim 0.1\,M_\odot$) was discovered to have a M dwarf companion orbiting inside its large disk cavity \citep{biller12,lacour16}. The presence of this companion provides a possible explanation for why a smaller, inner disk in this system appears to be highly misaligned \citep[$\sim$70\degr;][]{avenhaus14,marino15}, which may be driven by secular precession resonance between the disk and the companion \citep{owen17}.
The extreme mass ratio between the M dwarf and the primary ($q = 0.05$--0.10) puts the HD~142527 system in a different class of multiple star systems than those considered so far. Even though the system parameters do not map directly to sources like \obj, these dynamical effects are still important to consider, especially since most existing observations do not have sufficient sensitivity to detect such small companions for most sources (\obj\ included).

At younger ages, the Class I binary system L1551~IRS~5 is an interesting analogue because it hosts a large circumbinary disk \citep[$r\sim500$\,au, $M_\mathrm{disk} \approx 0.07 \,M_\odot$;][]{eisner12,takakuwa17} outside of two circumstellar disks. 
The binary is on a wide orbit ($a = 70\,\mathrm{au}$), to which the circumstellar disks are misaligned by up to $25^\circ$. 
The relative proper motion of the binary over a 15\,yr baseline indicates that it contains $1.7\,M_\odot$ of stellar mass on a 246\,yr orbit \citep{villa17}. 
The circumstellar disks are probably aligned with the rotation of the outer circumbinary envelope, which may indicate that the stars were formed by rotationally-driven fragmentation, preserving this orientation \citep{lim16}. 

The most recent example of a misaligned circumbinary disk is in the TWA~3A system \citep{kellog17}, which hosts a circumbinary disk within a hierarchical triple system of stars of near equal mass (SpTs M3--M4). The ``inner'' binary Aa-Ab has a 35 day eccentric ($e=0.63$) orbit and hosts a small disk extending 25 au in size \citep{andrews10}, while the ``outer'' orbit A-B takes 200--800 years. Although the absolute inclinations are not yet known, the parameter space is sufficiently constrained such that it is likely that all three planes (the inner orbit, outer orbit, and cirumbinary disk) are misaligned by at least $30^\circ$. The Aa-Ab circumbinary disk mutual inclination may be attributable to torques from the distant B companion.

Of all these T~Tauri sources surveyed, \obj\ stands out in terms of stellar mass. Its architecture proved relatively difficult to probe via traditional detection techniques, requiring sustained, long-term radial velocity monitoring over 35 years, as well as sophisticated care and attention to derive a radial velocity solution for the blended line profiles. Finally, resolved submm interferometric observations were necessary to measure the inclination of the circumtriple disk. Looking forward, sustained campaigns of astrometric monitoring will be most helpful for definitively constraining the orbital inclinations.

Although there may still be significant dynamical evolution in the architecture of the \obj\ system before it reaches the main sequence (MS), it is also worth briefly considering how its orbital parameters compare to the general MS population of triple stars, even though this population of older systems may have experienced significant dynamical evolution. Among late B star primaries (which \obj~A will be on the MS), 13\%\ of systems have multiplicity of three or higher \citep{eggleton08}; this fraction is roughly constant across spectral types B--G. In a detailed analysis of higher-order multiple systems, \citet{tokovinin97} found that the ratio $P_\mathrm{long} / P_{\rm short}$ was greater than 10 in almost all systems, presumably reflecting which orbits are stable. While \obj's ratio (17) is smaller than most, it is not an outlier among triple systems. \citet{tokovinin97,tokovinin17} find that the distribution of mutual inclinations between the orbital planes in triple systems is inconsistent both with complete alignment of the inner and outer orbits (zero mutual inclination) as well as completely independent inclinations (randomly distributed). For triples with outer projected separation $< 50\,$au the average misalignment is $20^\circ$, while orbits wider than 1000 au are not preferentially aligned. The RV + astrometric fits for \obj\ suggest that the stellar orbits are consistent with this picture (a mutual inclintion of $13 \pm 6$\degr). The population of misaligned triples may be the result of accretion of gas with randomly aligned angular momentum at the epoch of star formation.

Like most multiple stars, those in \obj\ likely formed through turbulent fragmentation of the molecular cloud, possibly at larger separations than they are now, and then hardened through decay via dynamical interactions, accretion, and the interaction of the circumstellar disks \citep{offner10,bate12}. Continued study of the \obj\ system, including spatially resolving the innermost regions to discover circumstellar disks and their relative inclinations, will be valuable to further understanding its formation process.

\section{Summary and Conclusions} \label{sec:summary}

In this work we:
\begin{itemize}
\item Used spatially and spectrally resolved ALMA observations of the \obj\ circum-triple disk to derive a dynamical mass of $M_\mathrm{tot} = 5.29 \pm 0.09\,M_\odot$. We find the disk is large, massive, and inclined at $i_\mathrm{disk} = 137.6\pm2.0$\degr.
\item Used 35 years of high resolution optical spectra to derive new radial velocities for both components in the inner binary. We then fit a hierarchical triple orbit and found a 241 day inner period (A-B) and an 11.5 year outer period (AB--C). When we combined the radial velocity constraints with the disk based constraint on $M_\mathrm{tot}$, we found stellar masses of $M_\mathrm{A}$=$2.7~M_\odot$, $M_\mathrm{B}$=$1.7~M_\odot$, and $M_\mathrm{C}$=$0.9~M_\odot$, to a precision of $\pm 0.3\,M_\odot$.
\item Combined the radial velocity data with the astrometric data from \citet{berger11} to perform a joint RV-astrometric fit and found large mutual inclinations between the stellar orbits and the disk ($\Phi_\mathrm{in} = 44\pm5^\circ$, $\Phi_\mathrm{out} = 54 \pm 7^\circ$). The stellar orbits may be mildly misaligned with each other ($\Phi_\mathrm{in/out} = 13 \pm 6^\circ$).
\item Placed \obj~A and B on the HR diagram, and found that their stellar properties are broadly consistent with the predictions of pre-main sequence models for their measured masses at an age of $\approx$1\,Myr.
\item Compiled a lightcurve with a 30 year baseline and identified several new eclipse events. We also identified a 0.2 mag amplitude mode of variability phased with the outer orbital period, which suggests the A-B binary may be partially obscured by micron-sized grains in the circumtriple disk cavity at outer apoastron.
\item Placed \obj\ in the context of other pre-main sequence multiple systems. While short period eccentric binary systems generally seem to have low mutual inclinations with their circumbinary disks, there are a number of longer period systems that exhibit significant mutual inclinations.
\end{itemize}

Given its uniquely large stellar mass, massive circumtriple disk, and puzzling eclipse behavior, \obj\ should remain a high priority target to study a unique class of dynamical interactions in pre-main sequence multiple systems.

\acknowledgments
IC acknowledges support from the Smithsonian Institution and Stanford KIPAC for the production of this manuscript. IC and EJ acknowledge useful conversations with Lisa Prato about \obj, and we greatly appreciate her sharing her results in advance of publication.  IC acknowledges helpful conversations with Maxwell Moe and Jack Lissauer about \obj\ and related systems; and would like to thank Eric Nielsen, Bruce Macintosh, and Rebekah Dawson for helpful discussions about astrometry and orbital dynamics. SA acknowledges the very helpful support provided by the NRAO Student Observing Support program related to the early development of this project. GT acknowledges partial support for this work from NSF grant AST-1509375. This paper makes use of the following ALMA data: 2012.1.00496.S. ALMA is a partnership of ESO (representing its member states), NSF (USA), and NINS (Japan), together with NRC (Canada) and NSC and ASIAA (Taiwan), in cooperation with the Republic of Chile.  The Joint ALMA Observatory is operated by ESO, AUI/NRAO, and NAOJ.  This research made extensive use of the Julia programming language \citep{bezanson17} and Astropy \citep{astropy13}.

\software{CASA \citep[v4.4;][]{mcmullin07}, IRAF \citep{tody86,tody93}, DiskJockey \citep{czekala15a}, RADMC-3D \citep{dullemond12}, emcee \citep{foreman-mackey13}, VARTOOLS \citep{hartman16}, Astropy \citep{astropy13}}

\bibliographystyle{yahapj.bst}
\bibliography{gwori.bib}

\end{document}